\documentclass[final]{siamltex}
\usepackage{url}

\usepackage[a4paper,margin=2.5cm]{geometry}
\usepackage[dvips]{graphicx}%for latex
\usepackage[dvips,usenames]{color}
\usepackage[dvips]{epsfig,colortbl}
\usepackage{amssymb,amsfonts, amsbsy,latexsym,graphics,setspace}
\usepackage{amsmath}
\usepackage{afterpage}
\usepackage{graphicx,subfigure,sidecap}
\usepackage[notref]{showkeys}
\newcommand{\paren}[1]{\left(#1\right)}
\newcommand{\D}[2]{\frac{d#1}{d#2}}
\newcommand{\PD}[2]{\frac{\partial#1}{\partial#2}}
\newcommand{\PDD}[3]{\frac{\partial^{#1}{#2}}{\partial{#3}^{#1}}}

\newcommand{\at}[2]{\left. #1 \right|_{#2}}

\newcommand{\abs}[1]{\lvert #1 \rvert}

\title{Asymptotic and bifurcation analysis of wave-pinning
in a reaction-diffusion model for cell polarization}
\author{Yoichiro Mori\thanks{
School of Mathematics, University of Minnesota,
Minneapolis MN 55455, USA.}
\and Alexandra Jilkine\thanks{
Green Center for Systems Biology \&
Department of Pharmacology,
University of Texas Southwestern Medical Center
Dallas TX 75390, USA.}
\and Leah Edelstein-Keshet
\thanks{
Institute of Applied Mathematics and Department of Mathematics.
University of British Columbia, Vancouver, B.C. Canada V6T 1Z2.}}

\begin{document}

\date{April 3, 2010}

\maketitle

\begin{abstract}
We describe and analyze a bistable reaction-diffusion (RD) 
model for two interconverting chemical species that exhibits
a phenomenon of \emph{wave-pinning}: a wave of activation of one of the
species is initiated at one end of the domain, moves into the domain,
decelerates, and eventually stops inside the domain, forming a stationary front.
The second (``inactive'') species is depleted in this process.
This behavior
arises in a model for chemical polarization of a cell 
by Rho GTPases in response
to stimulation. The initially spatially homogeneous concentration profile
(representative of a resting cell)
develops into an asymmetric stationary front profile
(typical of a polarized cell).
Wave-pinning here is based on three properties: (1) mass conservation in a finite
domain, (2) nonlinear reaction kinetics allowing for multiple
stable steady states,
and (3) a sufficiently large difference in diffusion of the two species.
Using matched asymptotic analysis, we explain the mathematical basis of
wave-pinning, and predict the speed and pinned position of
the wave.
An analysis of the bifurcation of the pinned front solution 
reveals how the wave-pinning regime depends on parameters
such as rates of diffusion and total mass of the species. 
We describe two ways in which the pinned solution can be lost depending 
on the details of the reaction kinetics:
a saddle-node or a pitchfork bifurcation.\\
\textbf{\em Manuscript submitted to SIAM Journal of Applied Mathematics, April, 4, 2010, under review.} 
\end{abstract}


\pagestyle{myheadings}
\thispagestyle{plain}

\markboth{Mori, Jilkine and Edelstein-Keshet}{Analysis of wave-based pattern formation mechanism}

\section{Introduction}

In a recent reaction-diffusion (RD) model for biochemical cell polarization proposed in \cite{mori_bj08}
we found a wave-based phenomenon whereby a traveling wave is
initiated at one end of a finite, homogeneous 1D domain, moves across the domain, but
stalls before arriving at the opposite end.
We refer to this behavior as {\it wave-pinning}.
We observed that this phenomenon was obtained from a two-component RD
system obeying a modest set of assumptions: (1) Mass is conserved
and limited, i.e. there is no production or removal, only exchange
between one species and the other.
(2) One species is far more mobile than the other, e.g. due to binding to immobile
structures, or embedding in a lipid membrane.
(3) There is feedback (autocatalysis) from one form to further
conversion to that form.

The biological motivation for studying our specific system
comes from internal chemical reorganization that is the initial
stage of \emph{polarization} of a living eukaryotic cell, 
such as a white-blood cell, amoeba, or yeast in response to a signal. 
Such  chemical asymmetry then organizes the downstream
response of the cell (e.g. shape change, motility, division, etc).
Explaining the basis for such symmetry breaking has
become an important question in cell biology over the past decade,
motivating such mathematical models as 
\cite{meinhardt_jcs_99,subramanian_jtb_04,narang_jtb_05,Otsuji-plos07,goryachev2008}.
Our own work has focused on the role of switch-like polarity proteins called Rho GTPases, which are
conserved in eukaryotic cells from amoebae to humans.
Upon stimulation, levels of Rho GTPase activity rapidly
redistribute across a cell.
% (Fig.~\ref{cell_view} \textbf{the current figure doesn't show this}). 
For example, some members of this family
(Rac, Cdc42) become strongly activated
at one end (which subsequently becomes the front of the cell \cite{kraynov_s_00,nalbant_s_04})
whereas others (such as RhoA) dominate at the opposite end (which becomes the rear
\cite{xu_c_03}).  Whereas in our
previous work we investigated such phenomena in the context of
the actin cytoskeleton and cell motion, \cite{maree_bmb_06,dawes_bj_07},
here we are concerned only with the mathematical basis for the initial symmetry breaking.
Originally, we explored multiple interacting Rho GTPases,
to determine how interactions between several members of this family
affect spatio-temporal dynamics \cite{maree_bmb_06,jilkine_bmb_07}.
In the more recent work \cite{mori_bj08}, we investigated a minimal
system, consisting of a single active-inactive pair of GTPases.
From a mathematical perspective, this yields an opportunity for  deeper
analysis. From a biological perspective, it clarifies what are minimal
conditions required for symmetry breaking.

The model described here and in \cite{mori_bj08} is consequently based on
the following abstraction of experimental observations about Rho proteins:
(1) The protein has an active (GTP-bound) and an inactive (GDP-bound) form.
(2) The active forms are exclusively found on the cell membrane;
those in the fluid interior of the cell (cytosol) are inactive.
(3) There is a 100-fold difference between rates of
diffusion of cytosolic vs membrane bound proteins \cite{postma_embo_04}.
(4) Continual active-inactive exchange is essential for proper polarization.
If this exchange is stopped, the cell cannot polarize \cite{irazoqui_ncb_03}.
(5) On the time-scale of polarization (minutes), there is little or no
protein synthesis in the cell (timescale of hours), i.e. during polarization,
the total amount of the given protein is roughly constant.
(6) Feedback from an active form to further activation are common.
A schematic diagram of our model is given in Fig.~\ref{cell_view}.
Cases where this
has been established experimentally include
\cite{wang_ncb_02,Kozubowski2008,park_mmbr_07}.

The RD system in this study
encapsulates all the above aspects, and
exhibits self-polarization as a result of the wave-like phenomenon described above.
The purpose of this paper is to investigate the properties of this model.
In Section~\ref{ModelFormul}, we formulate the model in one space dimension.
We first seek to obtain a mathematically clear picture of wave-pinning.
This is achieved by way of matched asymptotic calculations, presented 
in Section \ref{section_asymptotics}. Here, we shall see how the 
wave speed, shape and stall positions are affected by the parameters of the problem.
We also briefly discuss the behavior of higher dimensional generalizations
of the one-dimensional model. Next, we study the bifurcation structure 
of our system in \ref{section_bifurcation}. We delineate
the parameter regime for which wave-pinning is possible, and 
describe the bifurcation structures that are possible for different 
reaction kinetics.
A summary and biological implications are presented in the Discussion.

\section{Model formulation}
\label{ModelFormul}

\begin{figure}
\begin{centering}
\includegraphics[width=\textwidth]{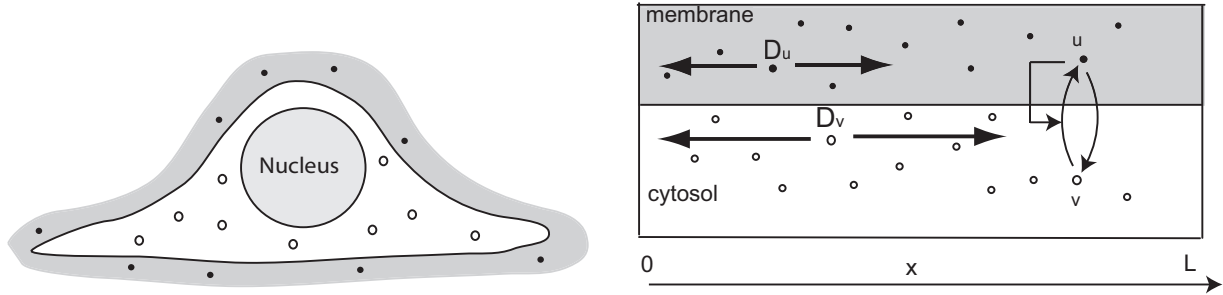}
\end{centering}
\caption{\footnotesize{
Left: The cell (initially unpolarized) consists of membrane (shaded) and cytosol (white).
The 1D model \eqref{OurModel} represents the chemical distribution 
of two forms of a protein $(u$, and $v)$ along the diameter of a cell,
idealized as a thin flat strip of uniform thickness. (The nucleus and other 
nonuniform features are neglected).
Right: in the RD system, $u(x,t)$ is an active protein, resident in and diffusing 
slowly along the membrane,
$v(x,t)$ is an inactive protein diffusing in the cytosol. Interconversion is 
subject to positive feedback from $u$ to itself. The cell diameter is along $0 \le x \le L$.
Cell and compartment sizes 
not drawn to scale.
}}
\label{cell_view}
\end{figure}

Consider a one dimensional domain $\Omega=\{x:0\leq x\leq L\}$.
Denote by $u(x,t)$ and $v(x,t)$ the concentrations of active and inactive
protein respectively at position $x$ and time $t$.
This one-dimensional model would be valid for a flat cell of 
sufficiently small thickness so
that appreciable chemical gradients do not develop
in the thickness direction (Fig. \ref{cell_view}). In this case, both membrane
and cytosolic positions can be described by a single coordinate $x$,
and thus, we may treat the membrane-bound species $u$ and the
cytosolic species $v$ as residing in the same domain $\Omega$.
There are biological situations that warrant models
in higher spatial dimensions, and we will consider such generalizations
in Section \ref{higher}.
From a mathematical point of view, however, we shall see that much of
the behavior of interest is already present in the
one-dimensional model.

The concentrations $u$ and $v$ satisfy the following equations
\begin{subequations}\label{OurModel}
\begin{align}
\PD{u}{t}&=D_u \PDD{2}{u}{x}+f(u,v),\\
\PD{v}{t}&=D_v \PDD{2}{v}{x}-f(u,v),
\end{align}
where $f(u,v)$ is the rate of interconversion of $v$ to $u$,
and the rates of diffusion satisfy  $D_u\ll D_v$, reflecting
the fact that the membrane bound species $u$ diffuses much more
slowly than the cytosolic species $v$.
The boundary conditions are
\begin{equation}\label{OurModelBCs}
\PD{u}{x}=\PD{v}{x}=0,\quad x=0,L.
\end{equation}
\end{subequations}
It is clear that system~\eqref{OurModel} leads
to mass conservation,
\begin{equation}\label{conservation}
\int_\Omega (u+v)dx=K_{\rm total},
\end{equation}
where $K_{\rm total}$ is a time-independent constant.

The following reaction term $f(u,v)$ was proposed in \cite{mori_bj08}:
\begin{equation}
\label{f Hill function}
f(u,v)= \eta \paren{\delta+\frac{\gamma u^2}{m^2+u^2}}v- \eta u
\end{equation}
where $\eta, \gamma, m>0, \delta\geq 0$ are constants. The above reaction term
is written as the difference between a production and a decay term.
The production term can be seen as $v$ times the production rate.
The production rate has a sigmoidal shape as a function of $u$,
which expresses the presence of positive feedback \cite{murray_93,tyson.cocb03}.
For suitable choices of $\gamma, m$ and $\delta$, $f(u,v)$ has the following property.
The expression $f(u,v)=0$, seen as an equation for $u$ with $v$ fixed
over a suitable range, has
three roots $u_-(v)<u_m(v)<u_+(v)$. Moreover, $u_{\pm}(v)$ are stable fixed
points of the ODE $\D{u}{t}=f(u,v)$ whereas $u_m(v)$ is an unstable fixed point.
In other words, the function $f(u,v)$ is a bistable function of $u$ over
a range of $v$ values.
Much of the analysis to follow applies not only to
the specific form of $f(u,v)$ given in \eqref{f Hill function}
but to a family of reaction terms satisfying a number of
properties including bistability.
A precise characterization of this family will be given shortly.

We now make our equations dimensionless.
We scale concentrations with $m$ and the reaction rate with
$\eta$, both of which are dictated by
the form of the reaction term (see \eqref{f Hill function}).
Take the domain length $L$ to be the relevant length scale.
Equations \eqref{OurModel} can be rescaled using
\begin{equation}\label{scaling}
u=m \tilde{u}, \quad v=m \tilde{v},
\quad x=L\tilde{x}, \quad t=\frac{L}{\sqrt{\eta D_u}}\tilde{t},
\end{equation}
where $\tilde{u}$, $\tilde{v}$, $\tilde{x}$, and $\tilde{t}$ are dimensionless variables.
The scaling in time is chosen so that we obtain a distinguished limit appropriate
for the analysis of wave-pinning (see next Section).
We define:
\begin{equation}\label{scalingparams}
\epsilon^2=\frac{D_u}{\eta L^2},\quad D=\frac{D_v}{\eta L^2}.
\end{equation}
Given $D_u\ll D_v$, we let $\epsilon$ be a small quantity.
We let $D=\mathcal{O}(1)$ with respect to $\epsilon$.
This assumption may be written as $\sqrt{D_v/\eta} \approx L$, i.e.
on the timescale of the biochemical reaction,
the inactive substance can diffuse across the domain.
In the context of cell polarization, we have a typical cell diameter $L\approx 10 \mu$m,
reaction timescale $\eta \approx 1\,$s$^{-1}$, and diffusion coefficients
$D_u=0.1 \, \mu$m$^2$s$^{-1}$ and $D_v=10 \, \mu$m$^2$s$^{-1}$.
The dimensionless constants are then $\epsilon \approx 0.03$ and $D\approx 0.1$.

Substituting the relationships \eqref{scaling} and \eqref{scalingparams} into
\eqref{OurModel} dropping the $\tilde{}$ and using the same symbol $f$ for
the dimensionless reaction term, we obtain:
\begin{subequations}\label{model}
\begin{align}
\epsilon\PD{u}{t}&=\epsilon^2\PDD{2}{u}{x}+f(u,v)\label{dlessU},\\
\epsilon\PD{v}{t}&=D\PDD{2}{v}{x}-f(u,v),\label{dlessV}
\end{align}
with boundary conditions:
\begin{equation}\label{no_flux_1D}
\PD{u}{x}=\PD{v}{x}=0, \quad  x=0,1.
\end{equation}
\end{subequations}
Note that our domain is now $0\leq x\leq 1$.
The reaction term \eqref{f Hill function} assumes the following dimensionless form:
\begin{equation}
f(u,v)=\paren{\delta +\frac{\gamma u^2}{1+u^2}}v-u. \label{hilldimless}
\end{equation}
The (dimensionless) total amount of protein satisfies
\begin{equation}\label{conservation_rescaled}
\int_0^1 (u+v)dx=K.
\end{equation}
where $K=K_{\rm total}/m$.
We shall henceforth work almost exclusively with the dimensionless system.

As mentioned earlier, we shall consider not only \eqref{hilldimless}
but a family of reaction terms satisfying the following properties:
\begin{enumerate}
\item (Bistability Condition) In some range $v_\text{min}\leq v \leq v_\text{max}$ ({\em bistable range}),
the equation $f(u,v)=0$ has three roots,
$u_-(v)<u_m(v)<u_+(v)$.  
Keeping $v$ fixed within the bistable range,  
$u_\pm(v)$ are stable fixed points
and $u_m(v)$ is an unstable fixed point of the ODE $\D{u}{t}=f(u,v).$
That is:
\begin{equation}
\PD{f}{u}(u_{\pm}(v),v)<0, \; \PD{f}{u}(u_m(v),v)>0.\label{bistability}
\end{equation}
\item (Homogeneous Stability Condition) The homogeneous states, $(u,v)\equiv (u_{\pm}(v),v), v_\text{min}<v<v_\text{max}$
are stable states of the system (\ref{model}).
\item (Velocity Sign Condition) There is one value $v=v_c, v_\text{min}<v_c<v_\text{max}$
at which the following integral $I(v)$ vanishes:
\begin{equation}
I(v)=\int_{u_-(v)}^{u_+(v)}f(u,v)\,du \label{I}.
\end{equation}
We assume in addition that $I>0$ for $v>v_c$ and $I<0$ for $v<v_c$.
\end{enumerate}
The first condition is the bistability condition that was mentioned earlier.
The reason for the name of the third condition will become clear in the next Section.
We shall see in Section \ref{stability} that the second condition can be reduced to
the following:
\begin{equation}\label{homstab}
\at{\paren{\PD{f}{u}-\PD{f}{v}}}{(u,v)=(u_{\pm}(v),v)}<0.
\end{equation}
Assuming this result,
we can check that \eqref{hilldimless} satisfies the above properties
for the following parameter values. For $\gamma>0$ and $\delta\geq 0$,
\eqref{hilldimless} satisfies the above conditions
if and only if
\begin{equation}
\gamma>8\delta.
\end{equation}
The corresponding 
bistable range is given by $v_\text{min}=\kappa_+<v<\kappa_-=v_\text{max}$ where:
\begin{equation}
\kappa_\pm=\frac{1}{\gamma}\paren{\frac{\rho}{\omega_\pm}+\frac{\omega_\pm}{1+\omega_\pm^2}}^{-1}, \;
\omega_\pm=\sqrt{\frac{1-2\rho\pm\sqrt{1-8\rho}}{2(1+\rho)}}, \; \rho=\frac{\delta}{\gamma}.
\end{equation}
When $\delta=0$, $v_\text{min}=2/\gamma $ and $v_\text{max}=\infty$. 
In our computational examples, we shall make use of \eqref{hilldimless} with $\delta=0$ and $\gamma=1$, 
which we record here for future reference:
\begin{equation}
f(u,v)=\frac{u^2v}{1+u^2}-u.\label{hill01}
\end{equation}
In this case, $u_0(v)$ and $u_\pm(v)$ can be computed explicitly:
\begin{equation}
u_-(v)=0, \; u_m(v)=\frac{v-\sqrt{v^2-4}}{2}, \; u_+(v)=\frac{v+\sqrt{v^2-4}}{2}.\label{hill01roots}
\end{equation}
We shall often make use of the following caricature of
\eqref{hilldimless}:
\begin{equation}\label{cubic}
f(u,v)=u(1-u)(u-1-v).
\end{equation}
It is easy to check that \eqref{cubic} satisfies all of the above properties.
For this reaction term, the bistable range is $0<v<\infty$.
This example makes certain algebraic manipulations easier than \eqref{hilldimless} or \eqref{hill01}.
In Section \ref{section_bifurcation}, we shall also make use of another
cubic that satisfies the above conditions:
\begin{equation}\label{cubicmod}
f(u,v)=-(u-1)(u-u_m)(u+1), \;\; u_m=-\frac{av}{\sqrt{1+(av)^2}},\;a>0.
\end{equation}
The bistable range for the model with kinetics ~\eqref{cubicmod} is $-\infty<v<\infty$.
At least one of the roots of this polynomial is always negative, and
thus, it is no longer possible to interpret $u$ and $v$ as being
concentrations of chemicals. The arguments
to follow, however, never require that $u$ and $v$ be positive.
Both \eqref{cubic} and \eqref{cubicmod} will prove useful in understanding the
bifurcation structure of our system.

\begin{figure}
\begin{centering}
\subfigure[]{\includegraphics[width=0.45\textwidth]{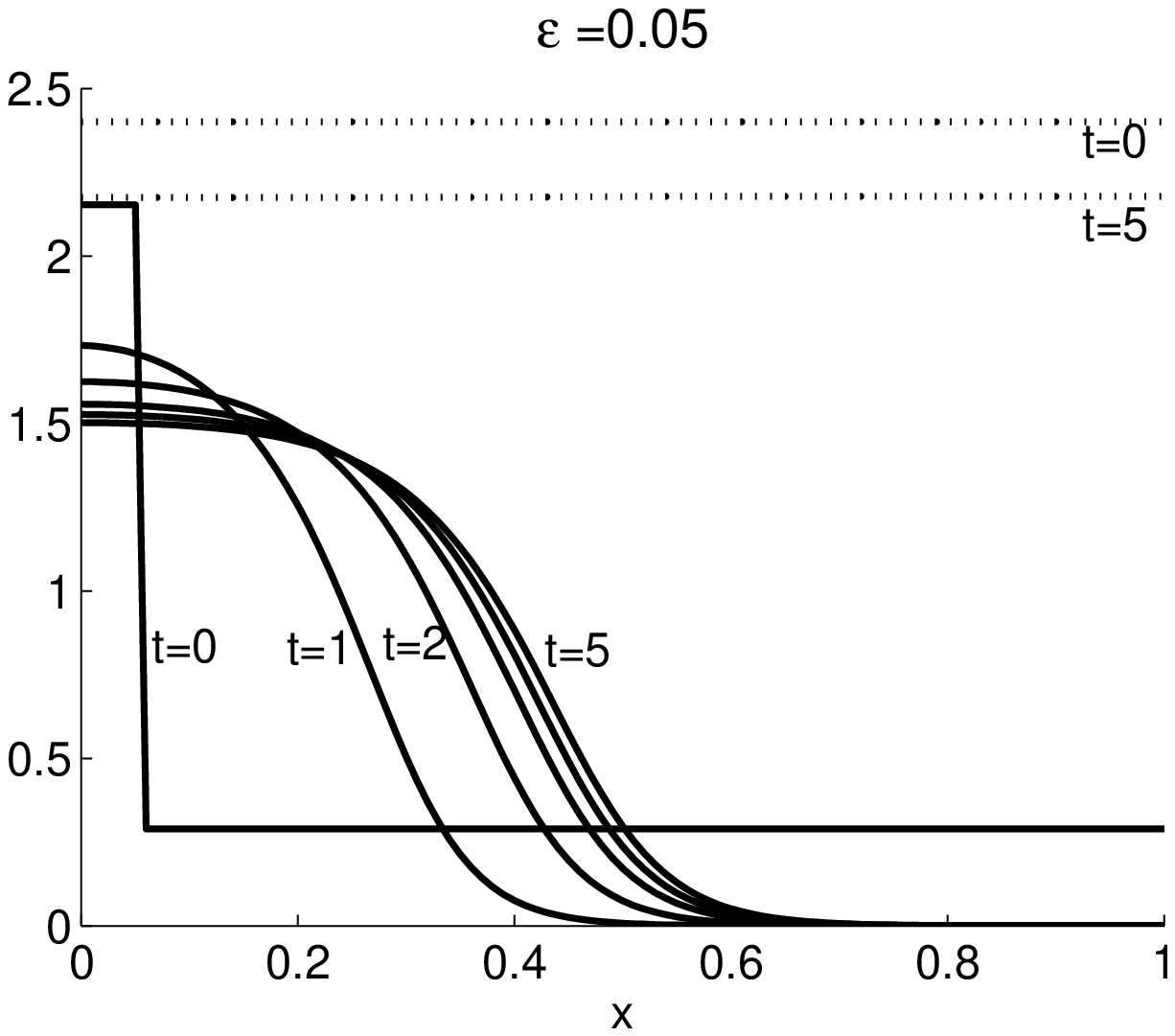}}
\subfigure[]{\includegraphics[width=0.45\textwidth]{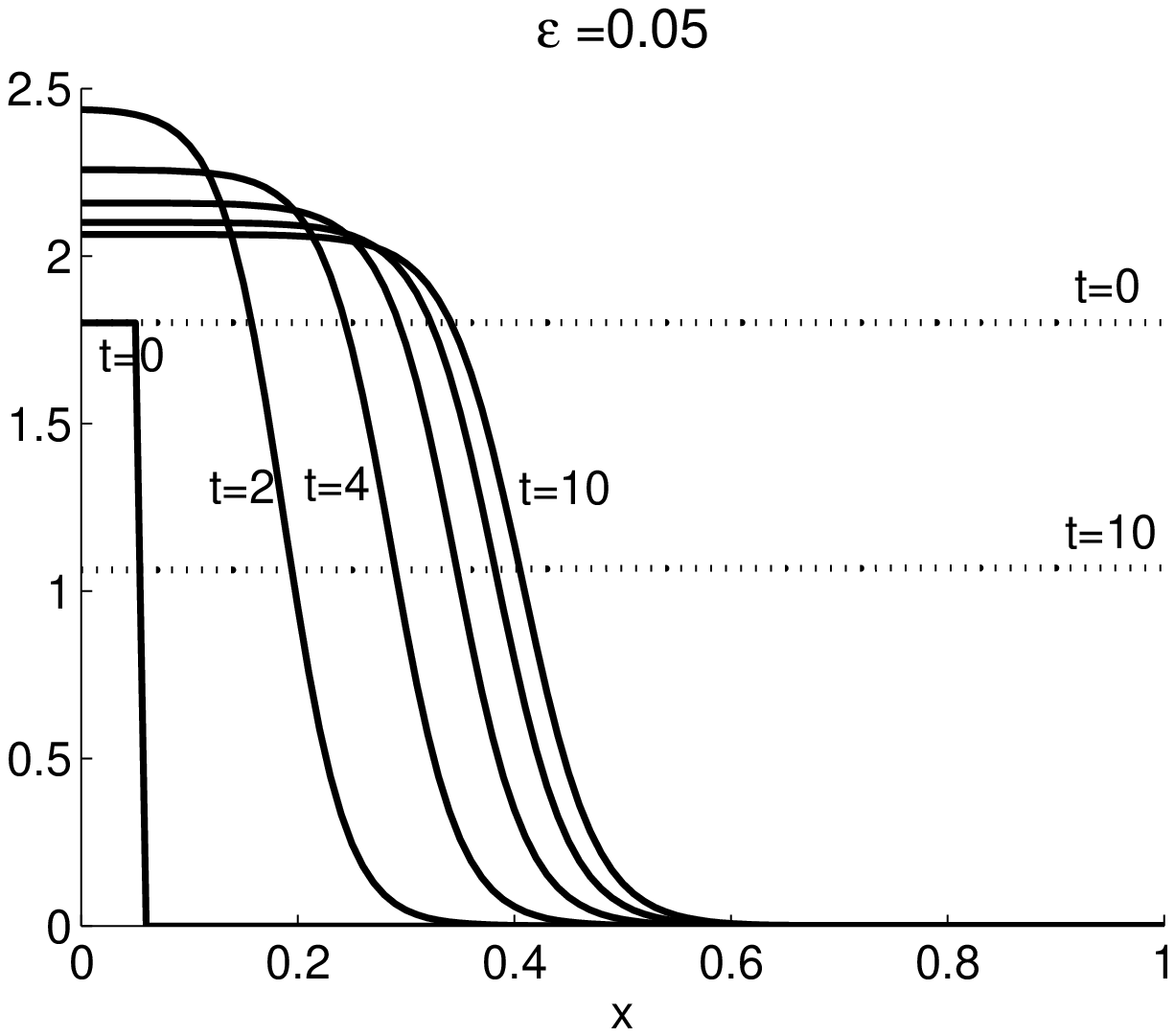}}
\par\end{centering}
\caption{\footnotesize{
Wave-pinning behavior for the reaction diffusion model \eqref{model} with
parameters $\epsilon=0.05$, $D=1$.
(a) Hill function reaction kinetics~\eqref{hill01}
with $\delta=0$, $\gamma=1$, $m=1$, $K=2.8$.
(b) Cubic  reaction kinetics~\eqref{cubic}
and $K=1.9$.
Solutions to $u$ (solid) and $v$ (dashed) are shown at the indicated times.
The wave is initiated as the square pulse in $u$, as shown at $t=0$.
}} \label{fig:wavepinExample}
\end{figure}

We now describe the behavior that we wish to explain.
If we consider \eqref{dlessU} as a stand-alone equation for fixed $v$,
it is a scalar reaction diffusion equation of bistable type. It is well-known
that such equations support propagating front solutions when posed on an
infinite domain. Coupling this with \eqref{dlessV} on a finite domain
gives rise to wave-pinning.
In Fig.~\ref{fig:wavepinExample}, we show simulation results for
our dimensionless system ~\eqref{model}
with the reaction terms \eqref{hill01} and \eqref{cubic}.
The concentrations are initialized so that $u$ is high
close to $x=0$ whereas $v$ is spatially uniform.
This represents a stimulus at the left end of the domain.
The initial rectangular profile of $u$ develops into a steep front
which propagates into the domain. The height of the front
gradually changes and the front eventually comes to a halt.
The left portion of the domain has a high concentration of the active
species $u$ whereas the right portion of the domain has a low concentration.
The spatially localized initial stimulus has been amplified to
produce a stable spatial segregation of the domain into a ``front''
and a ``back''. The one-dimensional cell has achieved polarization.

\section{Asymptotic Analysis of Wave-Pinning}\label{section_asymptotics}

In this Section, we perform an asymptotic analysis of wave-pinning
to obtain the speed of the wave and its stall position.
We shall first deal with the 1D RD system \eqref{model} and consider its higher dimensional
generalizations in Section \ref{higher}.

\subsection{Stability of the Homogeneous State}\label{stability}
Let $(u_s(v),v)$ be a steady state of \eqref{model},
where $u_s=u_{\pm}$ or $u_m$.
Linearize \eqref{model} about $(u_s,v)$:
\begin{equation}\label{linearized_sys}
\PD{}{t}
\begin{pmatrix}u\\v\end{pmatrix}
= \mathcal{L}
\begin{pmatrix}u\\v\end{pmatrix}
\equiv J
\begin{pmatrix}u\\v\end{pmatrix}
+\PDD{2}{}{x}
\begin{pmatrix}\epsilon^2 u\\Dv\end{pmatrix},
\quad \quad J=
\at{\begin{pmatrix}
f_u  & f_v\\
 -f_u & -f_v
\end{pmatrix}}{(u,v)=(u_s(v),v)}.
\end{equation}
where $f_u$ and $f_v$ denote partial derivatives of $f$
with respect to $u$ and $v$ respectively. Here,
the Jacobian of the reaction terms, $J$, is evaluated at $(u_s,v)$.
To study linear stability, we study the spectral properties of the
operator $\mathcal{L}$ under boundary conditions \eqref{no_flux_1D}.
We must also respect the mass constraint \eqref{conservation_rescaled}
so that the perturbations satisfy:
\begin{equation}
\int_0^1 (u+v)dx=0.\label{massless}
\end{equation}
Since we are on a bounded domain,
we need only consider eigenvalues.
We thus consider the eigenvalue problem:
\begin{equation}
\mathcal{L}\begin{pmatrix}u\\v\end{pmatrix}=
\lambda \begin{pmatrix}u\\v\end{pmatrix}.
\end{equation}
where $u,v$ must satisfy \eqref{no_flux_1D} as well as \eqref{massless}.
It is clear that all eigenfunctions are of the form:
\begin{equation}\label{linear_perturbation}
\begin{pmatrix}u\\v\end{pmatrix}
=\begin{pmatrix}\alpha_u\\\alpha_v\end{pmatrix}\cos kx
\end{equation}
where $k = n\pi$, $n=0,1,2,\cdots$ where $\alpha_u$ and $\alpha_v$ are constants
such that $(\alpha_u,\alpha_v)\neq (0,0)$.
When $k\neq 0$, $\alpha_u$ and $\alpha_v$ are arbitrary, whereas when $k=0$,
$\alpha_u+\alpha_v=0$ to satisfy \eqref{massless}.
Is is easily seen that the eigenvalues satisfy the quadratic equation:
\begin{equation}\label{quadratic}
\lambda^2-\tau_k \lambda +\Delta_k =0, \; \tau_k=\text{tr}\mathcal{L}_k,
\; \Delta_k=\text{det}\mathcal{L}_k, \;
\mathcal{L}_k=\begin{pmatrix}-\epsilon^2 k^2 + f_u & f_v\\ -f_u & -Dk^2-f_v \end{pmatrix}
\end{equation}
where $\text{tr}\mathcal{L}_k$ and $\text{det}\mathcal{L}_k$ denote, respectively,
the trace and determinant of the $2\times 2$ matrix $\mathcal{L}_k$.
Let us first consider the case $k=0$. In this case, the two solutions to
the above quadratic equation are:
\begin{equation}
\lambda=0, \; \text{ or }\lambda=\tau_0=f_u-f_v.
\end{equation}
If $\tau_0<0$, then the second eigenvalue is negative. As the reader
can easily check, as soon as
we assume $\tau_0=f_u-f_v\neq 0$,
the eigenfunction associated with $\lambda=0$ ceases
to satisfy the mass constraint. Thus, if we assume $\tau_0<0$ we have stability for $k=0$.
Now we turn to the case $k\neq 0$. Both roots of
\eqref{quadratic} have negative real part if and only if $\tau_k<0$ and $\Delta_k>0$.
Since
\begin{equation}
\tau_k=-(D+\epsilon^2)k^2+\tau_0<\tau_0
\end{equation}
$\tau_k<0$ so long as $\tau_0<0$.
\begin{equation}
\Delta_k=D\epsilon^2k^4-f_u(D-\epsilon^2)k^2-\tau_0\epsilon^2k^2\label{Delta}.
\end{equation}
Therefore, $\Delta_k>0$ so long as $f_u<0$ and $D>\epsilon^2$.
The condition $D>\epsilon^2$ is always met since we are assuming that $\epsilon$ is small.
For $u_s=u_{\pm}$, $f_u<0$ is met by the bistability condition \eqref{bistability}.
Thus, for $f$ satisfying the bistability condition, the homogeneous stability condition
of the last Section is equivalent to $\tau_0<0$. This is condition \eqref{homstab}.
It is interesting that the stability condition of the ODE system with fixed $v$ ($f_u<0$)
together with the stability condition for spatially homogeneous perturbations
($\tau_0<0$) implies stability for all wave numbers.

For $u_s=u_m$, $f_u>0$ by \eqref{bistability}.
For fixed $k$, \eqref{Delta} can be made negative by making $\epsilon$ sufficiently
small, and thus $(u_m(v),v)$ is always an unstable steady state for small enough $\epsilon$.
This does not preclude the possibility that $(u_m(v),v)$ be a stable steady state
for some finite $\epsilon$ value. Suppose $f_u>0$ and $\tau_0<0$. Let us consider
the positivity of $\Delta_k, k\geq \pi$:
\begin{equation}
\frac{\Delta_k}{k^2}=D\epsilon^2k^2-f_u(D-\epsilon^2)-\tau_0\epsilon^2
\geq D\epsilon^2 \pi^2-f_u(D-\epsilon^2)-\tau_0\epsilon^2.
\end{equation}
Therefore, $(u_m(v),v)$ is a stable steady state of the system so long as
the right-most quantity is positive. This is the case if
$\epsilon$ satisfies the following bound:
\begin{equation}
\epsilon^2>\frac{f_uD}{D\pi^2+f_v}.\label{epsbound}
\end{equation}
Since $\tau_0<0$ by assumption, $f_u<f_v$ and thus the right hand side of the
above inequality is less than $D$. Therefore, if $\tau_0<0$, there is a range of values
satisfying $\epsilon^2<D$ (i.e. diffusion coefficient of $u$ is smaller than that of $v$)
for which $(u_m(v),v)$ is a stable steady state of \eqref{model}.
On the other hand, if $\tau_0>0$, $(u_m(v),v)$ is always unstable.

For \eqref{cubic}, it is easily seen that $(u_m(v),v)$
is always unstable since $f_v=0$ at $u=u_m=1$ and thus $\tau_0=f_u>0$.
For \eqref{cubicmod}, $(u_m(v),v)$ can be stable for a range of $\epsilon$ values
if $a>1$ and $v$ is in a suitable range.
The stability of this middle stationary state does not play a role in the
wave-pinning analysis of the next Section. However, it does play a role in
determining the bifurcation structure of the system as we shall see in
Section \ref{otherbif}.

\subsection{Asymptotic Analysis of Wave-Pinning}\label{wavepinasymp}

We now consider the dynamics of \eqref{model}.
There are three time scales in this model, the short, intermediate
and long time scales. The intermediate time scale is of greatest interest to
us, and \eqref{model} is scaled accordingly. We shall
start with a brief discussion of the short time scale. Discussion of the long
time scale will be deferred to the next Section.

We introduce the short time variable $t_s=t/\epsilon$.
Then, \eqref{model} may be rewritten as:
\begin{subequations}\label{modelshort}
\begin{align}
\PD{u}{t_s}&=\epsilon^2\PDD{2}{u}{x}+f(u,v),\\
\PD{v}{t_s}&=D\PDD{2}{v}{x}-f(u,v).
\end{align}
\end{subequations}
with no-flux boundary conditions.
Assuming that $u$ admits an expansion in $\epsilon$ of the form $u=u_0+\epsilon u_1$
(and likewise for $v$), and substituting this into the above, we find that $u_0$ and $v_0$
satisfy the equations:
\begin{subequations}
\begin{align}
\PD{u_0}{t_s}&=f(u_0,v_0),\\
\PD{v_0}{t_s}&=D\PDD{2}{v_0}{x}-f(u_0,v_0).
\end{align}
\end{subequations}
Suppose $v_0$ satisfies $v_\text{min}<v_0<v_\text{max}$ so that $f(u_0,v_0)$ is bistable in $u_0$.
The first equation tells us that $u_0$ will evolve towards either $u_+(v_0)$ or $u_-(v_0)$
depending on whether $u_0(v_0)$ is greater or less than $u_m(v_0)$.
At the end of the short time scale, $v_0$ will have a spatial profile that is uniform
whereas $u_0$ will assume the values of $u_+(v_0)$ or $u_-(v_0)$ depending on position.
In other words, the domain will have segregated into regions where $u_0=u_+(v_0)$ or $u_-(v_0)$.
This profile will serve as our initial condition for the intermediate time scale.
In general, this initial profile consists of multiple transition layers where $u$ switches
its value from $u_+$ to $u_-$ or vice versa.
If there are no transition layers, this is nothing other than the stable steady state
whose stability we just studied.
In this Section, we shall restrict our
attention to the case when the initial profile consists only of a single transition layer.
Analysis in the case of multiple transition layers is essentially the same, and will
be discussed in the next Section.

We now begin the analysis in the intermediate time scale. Let $\phi(t)$ be the position
of the transition layer or the front. Note that the the position of the front changes
with time. We now perform a matched asymptotic calculation.

Expand $u=u_0+\epsilon u_1\cdots$ and likewise for $v$.
Substituting these expansions into
(\ref{model}a,b)
and retaining leading order terms we have the following equations for
$u_0$ and $v_0$:
\begin{subequations}\label{eq:1DouterLO}
\begin{align}
0&=f(u_0,v_0) \label{fU0V0},\\
0&=D\PDD{2}{v_0}{x}-f(u_0,v_0).
\end{align}
\end{subequations}
Equations \eqref{eq:1DouterLO} are valid in the outer region $0\leq x<\phi(t)-\mathcal{O}(\epsilon)$ and $\phi(t)+\mathcal{O}(\epsilon) \leq x<1$, that is, at some small distance
away from the sharp transition zone at the front.
Note that it is impossible to solve the above system with most initial
data for $u_0$ and $v_0$. This is the reason why we need to insert a short time scale
before this intermediate time scale. During the short time scale, the arbitrary initial
condition evolves into an initial profile that is admissible as an initial condition
for the intermediate time scale analysis.
Adding (\ref{eq:1DouterLO}a,b), we find:
\begin{equation}
D\PDD{2}{v_0}{x}=0.
\label{DV2dX2}
\end{equation}
From  \eqref{DV2dX2} and boundary conditions \eqref{no_flux_1D}, we conclude that:
\begin{equation}
v_0(x,t)=
\begin{cases}
v_<(t) & 0\leq x<\phi(t)-\mathcal{O}(\epsilon),\\
v_>(t) & \phi(t)+\mathcal{O}(\epsilon) < x \leq 1,
\end{cases}
\end{equation}
where the values of $v$ to the right and to the left of the front, $v_>$ and $v_<$, do not depend on $x$.
From (\ref{fU0V0}), $u_0$ takes on one of the values $u_+, u_-$ or
$u_m$
in the outer regions.
Since we are seeking a front solution, we let:
\begin{equation}\label{eq:1DOuterU0}
u_0(x,t)=
\begin{cases}
u_+(v_<) & 0\leq x<\phi(t)-\mathcal{O}(\epsilon),\\
u_-(v_>) & \phi(t)+\mathcal{O}(\epsilon) \leq x<1.
\end{cases}
\end{equation}
We have assumed, without loss of generality, that $u$ transitions from $u_+$ to $u_-$
in the direction of increasing $x$ as we traverse $\phi(t)$.

Let $w(x,t)=x-\phi(t)$
be the distance from the transition front, i.e. $w=0$ at the front.
Introduce a stretched coordinate $\xi$ for the inner layer  close to the evolving front:
\begin{equation}
\xi=\frac{w}{\epsilon}=\frac{x-\phi(t)}{\epsilon}.
\end{equation}
The inner solution is denoted by $U, V$, where
\begin{equation}
\label{eq:1DInnerUV}
U(\xi,t)=u((x-\phi(t))/\epsilon,t),\quad V(\xi,t)=v((x-\phi(t))/\epsilon,t).
\end{equation}
Note that \eqref{eq:1DInnerUV} is not a  traveling front solution in the strict sense,
as the wave speed $d\phi/dt$ is not constant.
As the amplitudes of $U$ and $V$ also change with time,
we do not assume $u(x,t)=U(\xi)$, but rather $u(x,t)=U(\xi,t)$, and likewise for $V$.

Substitute the new scaling into \eqref{model}
to obtain the inner equations on $-\infty<\xi<\infty$:
\begin{subequations}\label{eq:UVinner1D}
\begin{align}
\epsilon \PD{U}{t}-\D{\phi}{t}\PD{U}{\xi}&=\PDD{2}{U}{\xi}+f(U,V),\label{Uinner}\\
\epsilon \PD{V}{t}-\D{\phi}{t}\PD{V}{\xi}&=\frac{D}{\epsilon^{2}}\PDD{2}{V}{\xi}-f(U,V).\label{Vinner}
\end{align}
\end{subequations}
 Expanding
$U, V $ and $\phi$ in powers of $\epsilon$, we obtain, to leading order,
\begin{subequations}\label{eq:U0V01D}
\begin{align}
&\PDD{2}{U_0}{\xi}-\D{\phi_0}{t}\PD{U_0}{\xi}+f(U_0,V_0)=0, \label{U0_inner}\\
&\PDD{2}{V_0}{\xi}=0. \label{V0_inner}
\end{align}
\end{subequations}
From (\ref{V0_inner}), it follows that
\begin{equation}
V_0=a_1(t)\xi+a_2(t),
\end{equation}
where $a_1(t),\, a_2(t)$ are arbitrary functions of $t$ to be determined from matching.

\label{sec:matching1D}
We match the inner ($V_0$) and outer ($v_0$) solutions,
\begin{equation}
\lim_{\xi \rightarrow -\infty} V_0(\xi)=v_<,\quad \lim_{\xi \rightarrow \infty} V_0(\xi)=v_>.
\end{equation}
For these limits to exist,  $V_0$ must be
a constant in the inner layer, i.e.
\begin{equation}
v_0=V_0.
\end{equation} Thus, $V_0$
is spatially uniform throughout the domain, and is equal to the outer solution $v_0$.
We thus recover our observation that $v_0$ should be uniform by the time the dynamics
in the intermediate time scale dominates.
We drop the dependence of $v_0$ on $x$ (and $V_0$ on $\xi$).

We next consider a solution for $U_0$ in the inner layer. Since $V_0$ is spatially constant
in the inner layer,  \eqref{U0_inner} is an equation in $U_0$ only, where $V_0$ is
a parameter (that varies in time).
We must solve the boundary value problem (\ref{U0_inner}) with the matching conditions from \eqref{eq:1DOuterU0} as boundary conditions at $\pm \infty$:
\begin{equation}\label{matching_U0}
\lim_{\xi \rightarrow -\infty} U_0(\xi)=u_+(V_0),\quad \lim_{\xi \rightarrow \infty}U_0(\xi)=u_-(V_0).
\end{equation}
Such a heteroclinic solution $U_0^\phi(\xi,V_0)$, unique up to translation,
exists for general bistable reaction terms $f(U,V)$ \cite{keener_book,murray_93}.
Multiplying (\ref{U0_inner}) by ${\partial U_0^\phi}/{\partial \xi}$ and integrating from $\xi=-\infty$ to $\xi=\infty$,
we obtain:
\begin{equation}
\D{\phi_0}{t} \equiv c(V_0)= \frac{\int_{u_-(V_0)}^{u_+(V_0)}f(s,V_0)ds}
{\int_{-\infty}^\infty \paren{{\partial U_0^\phi(\xi,V_0)}/{\partial \xi}}^2d\xi}. \label{cV0}
\end{equation}
An explicit analytical expression for $c(v)$ cannot in general be obtained.
An exception is when the reaction kinetics is of the form
$f(u,v)=-(u-u_+(v))(u-u_m(v))(u-u_-(v))$, where $u_-<u_m<u_+$.
In this case $c(v)$ is given by \cite{murray_93}:
\begin{equation}
c(v)=\frac{1}{\sqrt{2}}\paren{u_+(v)-2u_m(v)+u_-(v)}. \label{Velocity_cubic}
\end{equation}
The sign of the velocity, however, is determined by the numerator of fraction in \eqref{cV0}
and can thus be easily determined given the reaction term $f(u,v)$.
By velocity sign condition (see equation \eqref{I}) we see that
$d\phi_0/dt$ is positive when $V_0>v_c$ and negative when $V_0<v_c$.

%% The unknown constant $V_0=v_0$ is obtained using a
%% solvability condition for the higher order outer equations.
%% Taking (\ref{model}a,b) to the next order
%% and isolating terms of order $\epsilon$ leads to
%% \begin{subequations}\label{eq:1DOuteruv}
%% \begin{align}
%% \PD{u_0}{t}=&f_u(u_0,v_0)u_1+f_v(u_0,v_0)v_1\label{u0t},\\
%% \PD{v_0}{t}=&D\PDD{2}{v_1}{x}-f_u(u_0,v_0)u_1-f_v(u_0,v_0)v_1.
%% \end{align}
%% \end{subequations}
%% Adding equations~\eqref{eq:1DOuteruv} we obtain
%% \begin{equation}\label{eq:1DOuteru0+v0}
%% \PD{}{t}(u_0+v_0)=D\PDD{2}{v_1}{x}.
%% \end{equation}
%% From \eqref{no_flux_1D}, we see
%% that $v_1$ satisfies no flux boundary conditions at $x=0$ and $x=1$.
%% Integrating \eqref{eq:1DOuteru0+v0} from $x=0$ to $x=1$ results in
%% \begin{equation} \label{eq:1DOuterInt(u0+v0)}
%% \PD{}{t}\paren{\int_0^1 u_0 \, dx+v_0}=0,
%% \end{equation}
%% where we used the
%% boundary conditions \eqref{no_flux_1D} and the fact that $v_0$ is a constant throughout the domain. Equation~\eqref{eq:1DOuterInt(u0+v0)} represents a solvability condition for equation~\eqref{eq:1DOuteru0+v0}. Integrating the above in $t$, leads to
%% \begin{equation}
%% v_0=K-\int_0^1 u_0 \,dx
%% \end{equation}
%% where $K$, as before, corresponds to the (dimensionless) total
%% amount of $u$ and $v$ in the domain.

By \eqref{conservation}, we see that $u_0$ and $V_0=v_0$ satisfy the relation:
\begin{equation}
v_0+\int_0^1 u_0 dx=K.
\end{equation}
The integral of $u_0$ can be approximated by contributions
from the two outer regions (to left and right of the front) and a $\mathcal{O}(\epsilon)$ 
contribution from the inner region:
\begin{equation}
\begin{split}
\int_0^1 u_0 \,dx &=\int_0^{\phi(t)-\mathcal{O}(\epsilon)}u_0\,dx+\int_{\phi(t)+\mathcal{O}(\epsilon)}^1 u_0\,dx
+\mathcal{O}(\epsilon) \nonumber \\
&=u_+(v_0)\phi_0(t)+u_-(v_0)(1-\phi_0(t))+\mathcal{O}(\epsilon),
\end{split}
\end{equation}
where we have used (\ref{eq:1DOuterU0}) in the second equality.
Discard terms of $\mathcal{O}(\epsilon)$.
The reaction-diffusion system is then reduced to
the following ordinary-differential-algebraic system:
\begin{equation}\label{V0}
\D{\phi_0}{t}=c(v_0), \quad
v_0=K- u_+(v_0)\phi_0-u_-(v_0)(1-\phi_0),
\end{equation}
where $c(v_0)$ is given by \eqref{cV0}.
In \eqref{V0}, the total amount of material, $K$, is allocated to
a band of width $\phi_0$ at level $u_{+}$,
a band of width $1-\phi_0$ at level $u_{-}$,
and a homogeneous level of $v_0$ across the entire interval.

We now show that the front speed, $d\phi_0/dt$, and the rate of change $dv_0/dt$, have opposite signs.
Differentiating the relation $f(u_\pm(v),v)=0$ with respect to $v$ and using \eqref{homstab} leads to
\begin{equation}
0=\at{\paren{\PD{f}{u}\D{u_{\pm}}{v}+\PD{f}{v}}}{u=u_\pm(v)}>\paren{1+\D{u_{\pm}}{v}}\at{\PD{f}{u}}{u=u_\pm(v)}.
\end{equation}
Using (\ref{bistability}) we conclude that:
\begin{equation}
1+\D{u_{\pm}}{v}>0. \label{oneplusdu}
\end{equation}
Differentiating the second relation in (\ref{V0}) with respect to $t$ results in:
\begin{equation}\label{dvdtanddphidt}
\paren{1+\D{u_+(v_0)}{v}\phi_0+\D{u_-(v_0)}{v}(1-\phi_0)}\D{v_0}{t}=-(u_+(v_0)-u_-(v_0))\D{\phi_0}{t}.
\end{equation}
Since the front position must reside
within a domain of unit length, we have $0<\phi_0<1$. Using this and (\ref{oneplusdu}),
we see that the factor multiplying $dv_0/dt$ in
(\ref{dvdtanddphidt}) is positive.
Since $(u_+-u_-)>0$, we conclude from (\ref{dvdtanddphidt})
that $dv_0/dt$ and $d\phi/dt$ have opposite signs. Thus, $v_0$ is depleted as the wave
progresses across the domain. It is interesting that this conclusion was
obtained using the two conditions, bistability and homogeneous stability.

Suppose $v$ is sufficiently large initially, i.e.,  $v_0>v_c$ at $t=0$.
Since $d\phi_0/dt$ is positive for $v_0>v_c$, $dv_0/dt<0$.
Thus, $v_0$ decreases as the front $\phi_0$ advances.
If $v_0$ approaches $v_c$ the front will come to a halt, i.e. will become pinned.
Suppose the front is pinned at $\phi_p$.
Then $\phi_p$ can be determined as follows.
When the wave pins, we have
\begin{equation}
v_c=K- u_+(v_c)\phi_p-u_-(v_c)(1-\phi_p).
\label{Vcandphieqn}
\end{equation}
We can interpret (\ref{Vcandphieqn})
as a relation between $\phi_p$ and $K$. We must have $0<\phi_p<1$.
This leads to a condition on $K$ for wave-pinning to occur:
\begin{equation}
v_c+u_-(v_c)<K<v_c+u_+(v_c) \label{pinconditiongeneric}
\end{equation}
that is, for wave-pinning to occur, the total concentration
of chemical in the domain must fall within a range given by
\eqref{pinconditiongeneric}.
The pinned front is stable; if the front is perturbed, it will relax back
to the pinned position $\phi_p$ as can be seen from the velocity sign condition
and the fact that $\D{\phi_0}{t}$ and $\D{v_0}{t}$ have opposite sign.

We now illustrate the above theory with the reaction term \eqref{cubic}.
In this case, the reaction term is a cubic polynomial in $u$,
and we may apply \eqref{Velocity_cubic} to find an explicit expression
for $c(v)$. The leading order equations \eqref{V0} become:
\begin{equation}\label{v0cubic}
\D{\phi_0}{t}=\frac{v_0-1}{\sqrt{2}}, \quad
v_0=K-(1+v_0)\phi_0.
\end{equation}
From \eqref{v0cubic}, we find that the wave stops when $v_0=1 \equiv v_c$.
Condition \eqref{pinconditiongeneric} reduces to:
\begin{equation}\label{1K3}
1<K<3.
\end{equation}

\begin{figure}[htbp]

\begin{centering}
\subfigure[]{\includegraphics[width=0.3\textwidth]{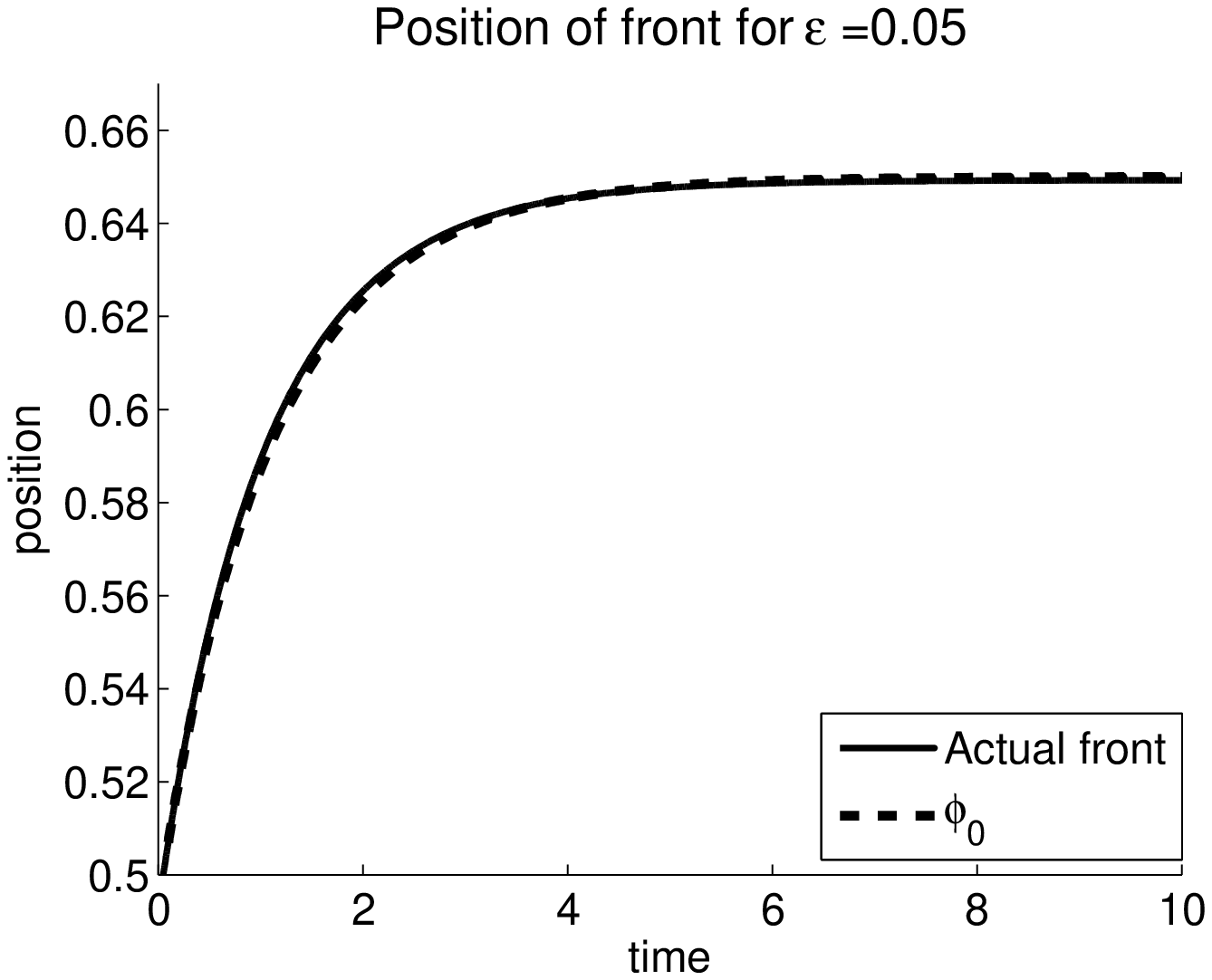}}
\subfigure[]{\includegraphics[width=0.3\textwidth]{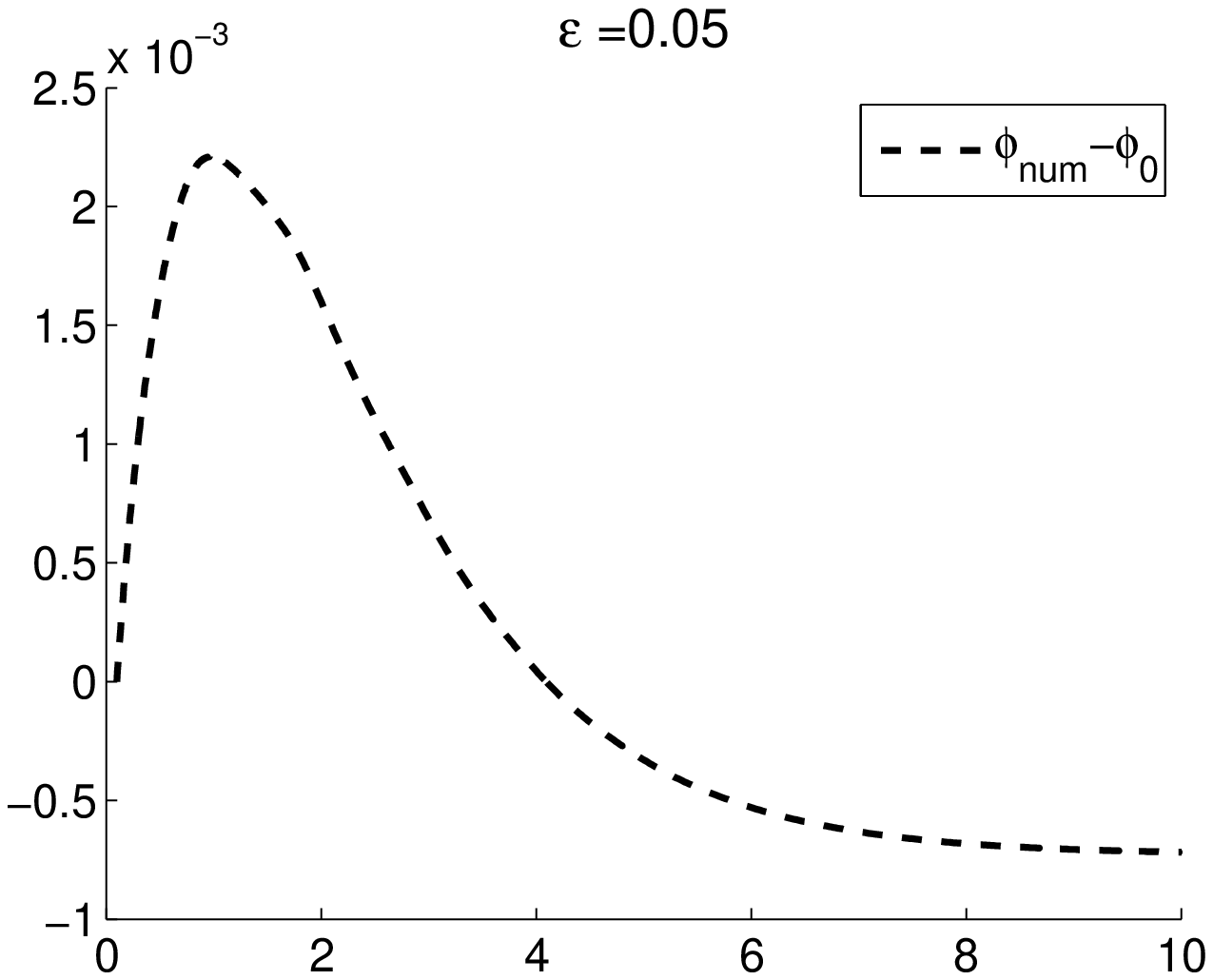}}
\subfigure[]{\includegraphics[width=0.3\textwidth]{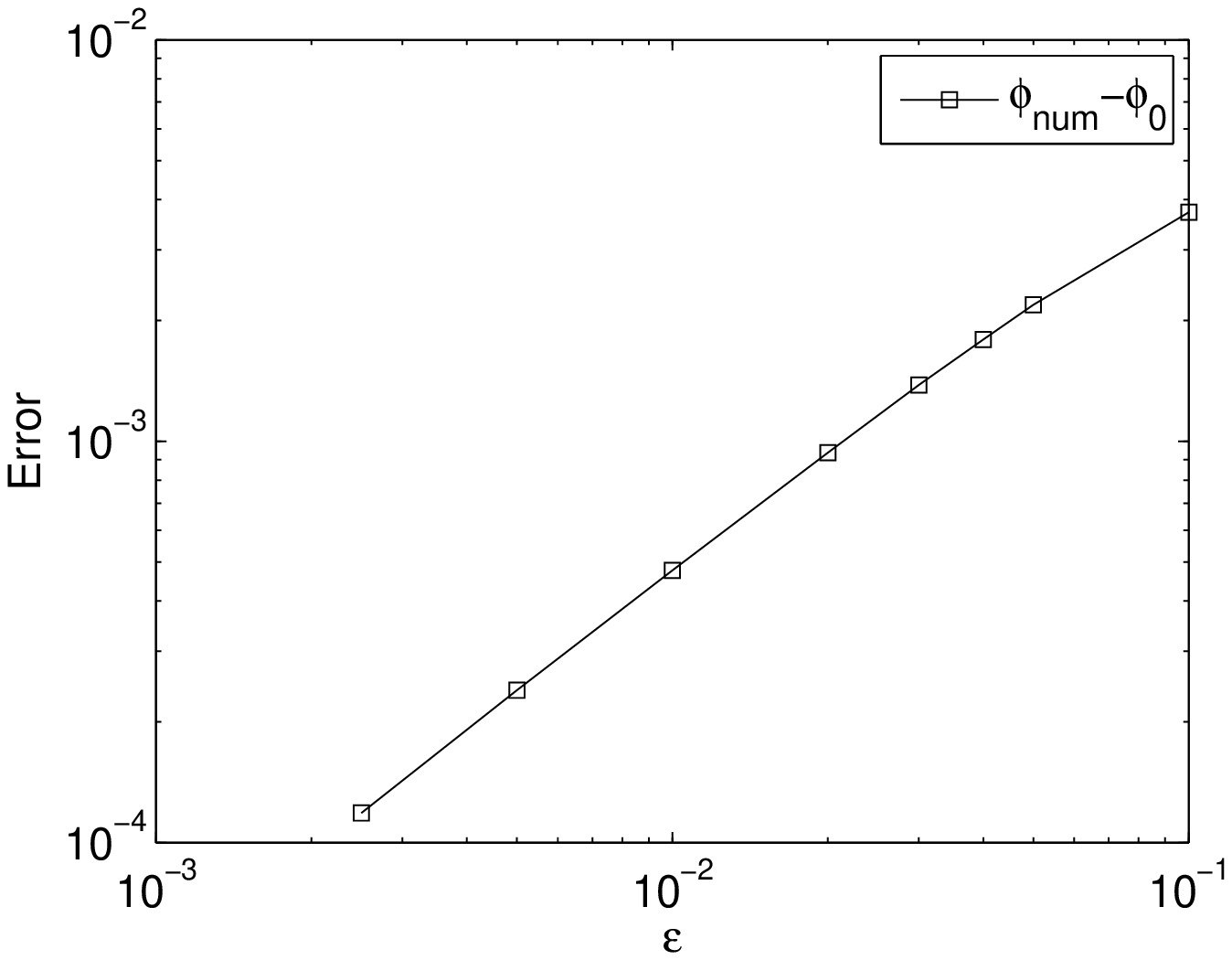}}
\par\end{centering}
\caption[The evolution of the steepest point
$\phi$ through time and the solution of Eq.~\ref{Velocity_example}]{\footnotesize{(a) The evolution of the front position
$\phi_\text{num}$  obtained by numerically solving the full PDE system \eqref{model}
with reaction kinetics \eqref{cubic} (solid), and the solution of the zero order asymptotic order approximation, $\phi_0$ from Eq.~\eqref{Velocity_example} (dashed). (b) The error $\phi_{num}-\phi_0 $ over time
%between$\phi_\text{num}$  and the solution of the zero order asymptotic order approximation (Eq.~\ref{Velocity_example}).
(c) The effect of $\epsilon$ on the error $\phi_{num}-\phi_0 $.
Parameters: $\epsilon=0.05$, $D=1$.}
%(c) The maximum error between the numerical front position $\phi_{num}$, and the leading order approximation \eqref{Velocity_example} is shown for a range of $\epsilon$ values.
} \label{Fig:Error_phi_0}
\end{figure}

Solving \eqref{v0cubic} for $v_0$, we obtain
 \begin{equation}\label{Velocity_example}
 \D{\phi_0}{t}=\frac{1}{\sqrt{2}}\paren{\frac{K-\phi}{1+\phi}-1}, \quad  v_0=\frac{K-\phi_0}{1+\phi_0}.
 \end{equation}
The position at which the wave stalls, is therefore
\begin{equation}\label{phi_p_example}
\phi_p = \frac{K-1}{2}.
\end{equation}
Fig.~\ref{Fig:Error_phi_0} shows that predictions of the ODE \eqref{Velocity_example}
agree with numerical solutions to the full PDE system \eqref{model} using the cubic reaction kinetics,
\eqref{cubic}.
The exact front position is calculated from the numerical solution of the PDE system
by tracking the position $\phi_\text{num}$ at which $u=u_m(v)$
($u_m=1$ for reaction kinetics \eqref{cubic}).
$\phi_\text{num}(t_0)$ is used as an initial condition,
where $t_0 \approx 0$ is a time at which the solution to the PDE system
has relaxed to the form assumed in the asymptotic calculations.
The error decreases with time as the wave becomes pinned.
Based on the numerical evidence, we find that the leading order approximation is accurate 
to order $\epsilon$.
To get a measure of the error of the leading term approximation, we can calculate the next term in the asymptotic expression. We refer the reader to \cite{jilkine_phdthesis}.

\subsection{Multiple Layers and Long Time Behavior}

In the previous section, we discussed the behavior of system \eqref{model} in the intermediate
time scale under the assumption that the initial profile consists only of a single front.
We discuss what happens when the initial profile has multiple fronts or layers.
Let $\phi_k(t), k=1,\cdots n$ be the front positions so that $\phi_k(t)<\phi_{k+1}(t)$.
For notational convenience, we let $\phi_0(t)=0$ and $\phi_{n+1}(t)=1$.
If $u$ transitions from $u_+$ to $u_-$ as we cross a front in the positive $x$ direction,
we shall call this a positive front. If the transition is from $u_-$ to $u_+$, we call
this a negative front. In the sequel, we shall assume that $\phi_1(t)$ is a positive front.
The case in which $\phi_1(t)$ is a negative front can be treated in an analogous fashion.
If $\phi_1(t)$ is a positive front, all fronts with odd $k$ are positive fronts
and all fronts with even $k$ are negative fronts. Through an analysis similar to
the one in the previous Section, we may conclude that
the dynamics of the fronts can be tracked by the following ODE system, similarly to
\eqref{V0}:
\begin{align}
\D{\phi_k}{t}&=c(v) \text{ if } 1\leq k\leq n \text{ is odd},\\
\D{\phi_k}{t}&=-c(v) \text{ if } 1\leq k\leq n \text{ is even},\\
K&=u_+(v)L_++u_-(1-L_+),\quad L_+=\sum_{0\leq 2l\leq n} (\phi_{2l+1}-\phi_{2l}).
\end{align}
For simplicity of notation, we have dropped the additional subscript
showing that the above are leading order approximations.
As the fronts evolve, it is possible that adjacent fronts will collide. In this
case, two fronts will disappear, and the dynamics can be continued by renaming
the fronts and applying the above ODE system with $n-2$ fronts instead of $n$ fronts.
If front $\phi_1(t)$ or $\phi_n(t)$ hits either $x=0$ or $x=1$ respectively,
one can again write down an ODE for the front positions with $n-1$ fronts valid
after this incident.

As $t\rightarrow \infty$ in the above ODE system, it is possible that
the final configuration will still consist of multiple fronts, despite
possible annihilations of fronts that may have occurred. At this point, $v=v_c$,
and all fronts have velocity $0$. As far as the intermediate time scale
is concerned, these multiple front solutions are stable.

A natural question is whether these multiple front solutions will slowly
evolve beyond the intermediate time scale.
In this long time scale, $v$ is almost exactly equal to $v_c$ everywhere. 
We shall not include a detailed analysis of the evolution of multiple
fronts in the long time scale, since the analysis
and results turn out to be very similar to that 
of the mass-constrained Allen Cahn
model, whose long time behavior has been studied extensively 
\cite{Ward_sjam_96,reyna1995metastable,sun2000dynamics}. 
More specifically,
the long time behavior of multiple front solutions of the equation:
\begin{equation}
\PD{u}{t}=\epsilon^2\PDD{2}{u}{x}+f(u,v_c)-\lambda,\quad \lambda=\int_0^1 f(u,v_c)dx
\end{equation}
with initial conditions satisfying:
\begin{equation}
v_c+\int_0^1 udx=K
\end{equation}
is the same to leading order
to the long time behavior of the multiple front solutions of our system.
In the mass-constrained Allen Cahn model, multiple front solutions are known
to slowly evolve to a single front solution. Thus, multiple front solutions
are metastable, and the only genuinely stable solutions are the single front
solutions. The time scale of this evolution is, however, ``exponentially slow'' \cite{sun2000dynamics}.

\subsection{Higher Dimensions}\label{higher}

We have thus far focused our attention on the 1D model. We may pose
similar equations in higher dimensions. Given a bounded domain in $\Omega$ in $\mathbb{R}^n$,
we may pose the following problem:
\begin{subequations}\label{hdmodel}
\begin{align}
\epsilon\PD{u}{t}&=\epsilon^2\Delta u+f(u,v),\\
\epsilon\PD{v}{t}&=D\Delta v-f(u,v),
\end{align}
\end{subequations}
with no-flux boundary conditions on the boundary $\partial \Omega$.
If $n=2$, we may view this model as modeling a top-down view of a cell of (small) uniform thickness. Given that $u$ is a membrane
bound species, it may be more appropriate in the cell biological context to study the following.
Let $\Omega$ be a smooth domain in $\mathbb{R}^3$ whose boundary is the cell membrane $\Gamma$.
The membrane bound species $u$ resides entirely on the membrane $\Gamma$ whereas $v$ diffuses
freely inside the cell. We may write down the following equations:
\begin{subequations}\label{hdmemmodel}
\begin{align}
\epsilon\PD{u}{t}&=\epsilon^2\Delta_\Gamma u+f(u,v) \text{ on } \Gamma,\\
\epsilon\PD{v}{t}&=D\Delta v \text{ in } \Omega,\\
-D\PD{v}{\mathbf{n}}&=f(u,v) \text{ on } \Gamma.
\end{align}
\end{subequations}
where $\Delta_\Gamma$ is the Laplace-Beltrami operator associated with the surface $\Gamma$
and $\PD{v}{\mathbf{n}}$ is the normal derivative of $v$ taken in the outward direction.
It is easily seen that:
\begin{equation}
\D{}{t}\paren{\int_\Gamma udx+\int_\Omega v dx}=0
\end{equation}
where $\int_\Gamma \cdot dx$ denotes surface integration over $\Gamma$.

We now include a brief discussion of the behavior of these systems.
We shall first focus on system \eqref{hdmodel} and comment on \eqref{hdmemmodel} later.
In the short time scale, the domain $\Omega$ will be segregated into a portion $\Omega_+$
where $u\sim u_+$ and $\Omega_-$ where $u\sim u_-$. The concentration $v$ is uniform
throughout the domain.  This serves as the initial profile
for the intermediate time scale. The dynamics in the intermediate time scale may
be analyzed analogously to what was done for the 1D case. Introduce a stretched coordinate
near the transition layers, and perform matched asymptotics. This analysis is slightly
complicated by the fact that the transition layers lie along curves if $\Omega\subset \mathbb{R}^2$
(or hypersurfaces if $\Omega\subset \mathbb{R}^n$) and thus, requires the introduction of
curvilinear coordinates.
However, to leading order in the intermediate time scale,
effects due to the geometry of the transition layers turn out to be higher order.
Thus, the fronts advance with the normal velocity equal to $c(v)$ where $v$ is spatially uniform
and determined by mass conservation.

The dynamics in the long time scale reduces to the dynamics of the mass-constrained Allen Cahn model.
The transition layers evolve by mean curvature, with the constraint that the area (or volume)
of $\Omega_+$ must be conserved. We refer to \cite{rubinstein_imajam_92,Ward_sjam_96} 
for further details of the long time behavior
of the mass-constrained Allen Cahn model.

The dynamics of \eqref{hdmemmodel} is similar. In the short time scale, $\Gamma$ segregates into
domains $\Gamma_+$ and $\Gamma_-$ where the concentration of $u$ is $u_+$ and $u_-$ respectively,
whereas $v$ is uniform throughout $\Omega$. In the intermediate time scale, the transition curves
that reside on $\Gamma$ advance so that the normal velocity is equal to $c(v)$. In the
long time scale, we expect that the transition layers will evolve by {\em geodesic}
curvature with the constraint that the area $\Gamma_+$ must be conserved.

Although this long time behavior is interesting from a mathematical point of view,
it is not clear whether this has any relevance in the cell biological setting since
many other biochemical pathways will have come into the picture by that time.

\section{Bifurcation Structure of the Wave-Pinning System}
\label{section_bifurcation}

\begin{figure}
\begin{centering}
\includegraphics[width=\textwidth]{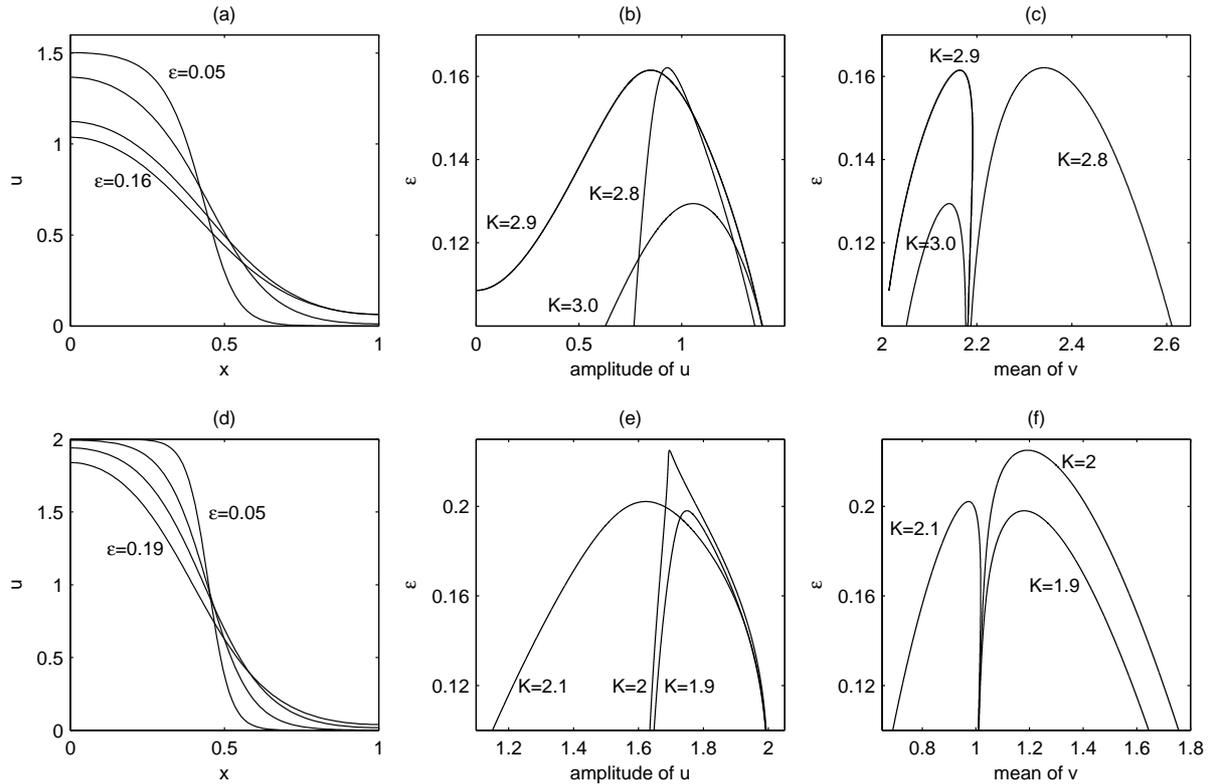}
\end{centering}
\caption{\footnotesize{
%Sample computations as $\epsilon$ is increased.
The effect of $\epsilon$ on wave shape and existence/stability for
reaction terms  \eqref{hill01} (a-c) and \eqref{cubic} (d-f) with $D=1$.
%In (a), (b), (c), \eqref{hill01} is used whereas \eqref{cubic} is used for (d), (e), (f).
(a,d): the shape of the pinned wave with (a) $K=2.8$ for $\epsilon=0.05,0.1,0.15,0.16$ and 
(d) $K=1.9$ for $\epsilon=0.05,0.1,0.15,0.19$.
The front gets shallower and broader as $\epsilon$ increases, losing stability at  
%and becomes unstable at 
%The pinned front loses stability at about 
(a) $\epsilon_c\approx 0.1621$ %for (a) and 
(d) $\epsilon_c\approx 0.1980$.
% for (d).
% In (b, e) and (c, f) 
 We plot the amplitude of $u$ (in b, e) and the mean of $v$  (in c, f) as
the pinned solution is continued for $K=2.8,2.9,3.0$ for (b, c) and $K=1.9,2,2.1$ for (e, f).
The peaks correspond to saddle-node bifurcations. In (b), the amplitude approaches
$u_+(v_c)\approx 1.5150$ for the pinned front and decreases as the solution is continued.
For $K=2.9$, the amplitude reaches $0$, at which there is a pitchfork bifurcation 
(see Section \ref{otherbif}). In (c), the mean of $v$ is close to $v=v_c=2.17506$ for the pinned solution.
In (e), the amplitude
is close to $2$ for the pinned front.
In (f), the mean of $v$ is close to $1$ for the pinned solution. 
Note the similarity of (f) with Fig. \ref{bifdiagram}.
}}
\label{samplebif}
\end{figure}

As we saw in the previous section, wave-pinning behavior
occurs for small values $\epsilon$. In this regime, the pinned
single front solution (which we shall hence forth refer to as the {\em pinned solution} or {\em pinned front})
is a stable stationary solution of the system.
A natural question is whether this pinned solution persists
as the value of $\epsilon$ is increased. If $\epsilon^2=D$ (or $\epsilon=\sqrt{D}$) in \eqref{model}
(the diffusion coefficient of the two species are the same),
such stable front solutions cannot exist. We thus expect that there
is some value of $\epsilon$
above which the pinned solution ceases to exist.
We simulated \eqref{model} to steady state
and gradually increased the value of $\epsilon$.
Fig. \ref{samplebif}(a) and (d) show the results of
sample computations when \eqref{hill01} and \eqref{cubic} 
are used for the reaction term.
For small $\epsilon$ values,
there is a gradual change in the wave shape and stall position.
Beyond some $\epsilon_c$, the pinned front disappears and is replaced
by a stable spatially homogeneous solution.
An interesting feature of this transition is that it is ``abrupt'':
the amplitude of the front (the
difference between the maximum and minimum values of $u$) does not vanish
gradually as $\epsilon$ approaches $\epsilon_c$. In Section \ref{finiteD}, we shall 
explore the bifurcation structure for \eqref{hill01} and \eqref{cubic}.
In Section \ref{secDinfty}, we focus on obtaining detailed information
on the bifurcation structure for \eqref{cubic}. 
In Section \ref{otherbif}, we indicate
other possible bifurcations we may expect of the pinned solution.

\subsection{Bifurcation at finite $D$}\label{finiteD}

For fixed $D>0$ and $K$ chosen in a suitable range,
there is a stable front solution to system
\eqref{model} (i.e., the pinned front) for $\epsilon$ sufficiently small.
As we just saw, there is a value $\epsilon=\epsilon_c$ above which
this pinned solution can no longer be continued.
The value $\epsilon_c$ is thus a function of $D$ and $K$.

To compute $\epsilon_c$, we first note that
stationary solutions satisfy the following system of equations:
\begin{subequations}\label{steadystate}
\begin{align}
0&=\epsilon^2 \PDD{2}{u}{x}+  f(u,v), \label{u_steady_state}\\
0&=D \PDD{2}{v}{x}-  f(u,v), \label{v_steady_state}\\
\PD{u}{x}&=\PD{v}{x}=0 \text{ at } x=0,1,\label{bc}\\
K&=\int_0^1 (u+v)dx.\label{uvK}
\end{align}
\end{subequations}
For fixed $D$ and $K$, we may numerically compute
$\epsilon_c$ by the following
procedure. First, set $\epsilon=\epsilon_0$ sufficiently small
so that the model \eqref{model} supports wave-pinning behavior.
We discretize \eqref{steadystate} and use a Newton iteration to find the
stationary solution, first for $\epsilon=\epsilon_0$.
The solution to the above equations is far from unique, and we
thus need a good initial guess to obtain the stationary solution
of interest to us, the pinned solution.
In order to do so, we simulate the time-dependent
model \eqref{model} to steady state (i.e., run the simulation
long enough) with initial conditions
suitable to produce a pinned solution as $t\rightarrow \infty$.
This computed steady state serves as our initial
guess to find the steady state solution at $\epsilon_0$.

Once we have the pinned solution at $\epsilon_0$, we may
increase $\epsilon$ to $\epsilon=\epsilon_1>\epsilon_0$ and use
the pinned solution at $\epsilon_0$ as our starting point
for the Newton iteration. The stability of the newly obtained
stationary solution can be monitored by computing the spectrum
of the discretization of the linearized operator $\mathcal{L}$
around the computed stationary solution.
This process can be repeated to find $\epsilon_c$,
the point above which a front solution can no longer be obtained.

Determination of $\epsilon_c$ using the above procedure,
however, is unreliable since the Jacobian matrix for the
Newton iteration becomes increasingly singular as $\epsilon$ approaches
$\epsilon_c$.
We could not obtain values of $\epsilon_c$ beyond
an accuracy of approximately $1.0\times 10^{-3}$.
A well-known remedy for this is to use pseudoarclength continuation
\cite{seydel_book}.
Instead of using $\epsilon$ as the bifurcation parameter, one uses
a {\em pseudoarclength} variable along the bifurcation curve so that
we can successfully continue the solution across a fold singularity.
This also gives us information on the bifurcation structure
at $\epsilon=\epsilon_c$.

Results from such a computation are given  
in Fig. \ref{samplebif} (panels b, c for
reaction terms \eqref{hill01} and panels e, f for reaction terms \eqref{cubic}).
For all values of $D$ and $K$ tested,
the numerical results indicated a fold (saddle-node)
bifurcation at $\epsilon_c$. The pinned solution is stable until
$\epsilon=\epsilon_c$ is reached, and this merges with a front
solution which has a one-dimensional unstable direction.

\begin{SCfigure}
\begin{centering}
\includegraphics[width=0.7\textwidth]{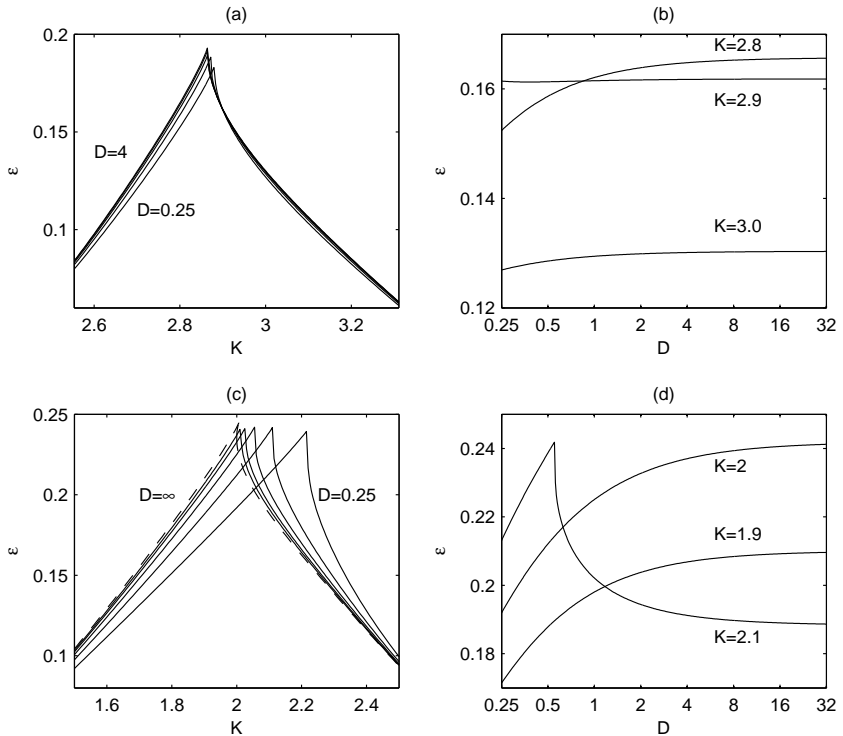}
\end{centering}
\caption{\footnotesize{
Composite families of two-parameter bifurcation diagrams showing the wave-pinning
regimes (always below the displayed curve(s)) in the $K\epsilon$ and $D \epsilon$ planes
for kinetics \eqref{hill01} (in a, b) and \eqref{cubic} (in c,  d). 
%
%A plot of $\epsilon_c$ as a function of $D$ and $K$. 
%Reaction term \eqref{hill01} is used in figures (a), (b) and \eqref{cubic}
%in figures (c), (d). 
In (a) and (c), the critical value of $\epsilon$, $\epsilon_c$, is plotted as a function of $K$ for fixed
$D$. The five solid lines correspond to $D=0.25,0.5,1,2,4$.
The dashed line in (c) is the
$D\rightarrow\infty$ curve (computed separately, see Section \ref{secDinfty}).
In (b) and (d), $\epsilon_c$ is plotted as a function of $D$ for fixed
$K$ where $K=2.8,2.9,3.0$ in (b) and $K=1.9,2.0,2.1$ in (d). 
The $D$-axes are scaled logarithmically.}}
\label{epsDfinite}
\end{SCfigure}

The dependence of $\epsilon_c$ on $D$ and $K$
is plotted in Figure \ref{epsDfinite}. Recall from \eqref{1K3} that
$K_\text{min}=1<K<3=K_\text{max}$ is the wave-pinning regime for \eqref{cubic}. 
For \eqref{hill01}, the wave-pinning regime is given by
\eqref{pinconditiongeneric} where $u_\pm$ is given in \eqref{hill01roots}
and $v_c\approx 2.17506$ is the solution to the equation (see \eqref{I}): 
\begin{equation}
I(v_c)=\int_{u_-(v_c)}^{u_+(v_c)}f(u,v_c)du=v_c(u_+(v_c)-\arctan(u_+(v_c)))-\frac{1}{2}(u_+(v_c))^2=0.
\end{equation}
The wave-pinning regime is $K_\text{min}<K<K_\text{max}$ where $K_\text{min}=v_c, 
K_\text{max}\approx 3.6901$. For both \eqref{hill01} and \eqref{cubic}, 
we sampled $K$ between $(3K_\text{min}+K_\text{max})/4<K<(K_\text{min}+3K_\text{max})/4$.

For both reaction terms, if we fix $D$, we see that there is a value $K=K_m$ at which
$\epsilon_c$ reaches a sharp peak. The value of $K_m$ is located roughly 
at $(K_\text{min}+K_\text{max})/2$.
This may be rationalized as follows.
The asymptotic calculations were based on the front being
sharp and away from the boundaries. When $K=(K_\text{min}+K_\text{max})/2$, the front is
positioned approximately in the middle of the domain,
and thus farthest from the boundaries.
One may thus argue that $K=(K_\text{min}+K_\text{max})/2$ would tend to
``maximize'' the range of validity of the asymptotic calculations.
The reason for the peaked appearance of the $\epsilon_c$
plot for fixed $D$ will be addressed toward the end of the next section.

The computed results indicate that $\epsilon_c$ is uniformly bounded
in $D$ and $K$. In particular,
we observe that, for fixed $K$,
$\epsilon_c(D,K)$ tends to some value as $D$ is taken large.
This serves as one motivation for studying the limit $D\rightarrow \infty$.

\subsection{Bifurcation Diagram in the limit $D \to \infty$}\label{secDinfty}

Here, we study the bifurcation structure of the following system:
\begin{subequations}\label{Dinfty}
\begin{align}
\epsilon\PD{u}{t}&=\epsilon^2\PDD{2}{u}{x}+f(u,v),\\
v&=K-\int_0^1 udx,
\end{align}
\end{subequations}
where $u$ satisfies no-flux boundary conditions at $x=0$ and $x=1$.
The above system should be seen as the limiting system of \eqref{model}
when $D\rightarrow \infty$. In this limit, $v$ is spatially
uniform and hence the evolution of $v$ is determined completely by
the mass constraint. By retracing the arguments of Section \ref{wavepinasymp},
the reader can easily check that the above system exhibits the
same wave-pinning behavior as \eqref{model}. In fact, the analysis
is somewhat simpler. When $D$ is finite, we deduce from our
asymptotic ansatz that $v$ is spatially uniform to leading order.
In the case of \eqref{Dinfty}, the spatial uniformity of $v$ is
automatic. We note that system \eqref{Dinfty} is often referred to as
the {\em shadow system} of \eqref{model}, and that it has been shown to 
provide insight into the behavior of the original system 
\cite{nishiura1982global,fusco1989slow,hale1989shadow}.

In studying the bifurcation of system \eqref{Dinfty}, we may
proceed similarly to the finite $D$ case of the previous section.
However, we shall take a different approach that will allow us
to obtain a far more detailed picture of the bifurcation structure.

\begin{SCfigure}
\begin{centering}
\includegraphics[width=0.7\textwidth]{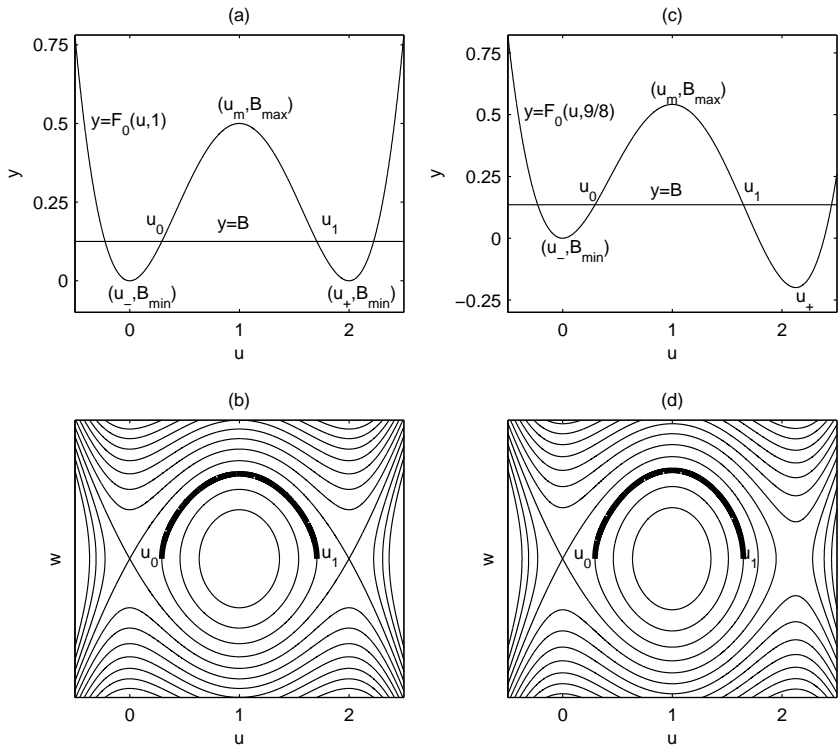}
\end{centering}
\caption{\footnotesize{
Typical shapes of the functions $y=F_0(u,v)$ (panels a,c)
and level curves of $w^2=F_0(u,v)$ (panels b, d) in the 
$uw$ phase planes for kinetics \eqref{cubic} for $v=1$ (right)
and $v=9/8$ (left).
%Two sample $u-w$ phase planes when \eqref{cubic} is used for the reaction term. On panel (a), we plot the functions $y=F_0(u,v), v=1$ with $y=B$. On panel (b), we plot the level curves of $w^2=F_0(u,v)$. 
It is clear that
there can only be a half-loop trajectory when $F_0(u,v)=B$ has four
distinct solutions. This happens when $B_\text{min}<B<B_\text{max}$.
%On panel (c) and (d) we have a phase plane example for $v=9/8$.
Note that, as $B\searrow B_\text{min}$, the half loop
approaches either a heteroclinic or (half of) a homoclinic orbit.
As $B\nearrow B_\text{max}$, the half loop approaches the neutral center
$(u,w)=(u_m,0)$.}}
\label{phaseplane}
\end{SCfigure}

The steady state solutions to \eqref{Dinfty} satisfy:
\begin{subequations}\label{Dinftysteady}
\begin{align}
0&=\epsilon^2\PDD{2}{u}{x}+f(u,v),\\
\PD{u}{x}&=0 \text{ at } x=0, 1,\label{nofluxux}\\
v&=K-\int_0^1 udx.\label{uvKinf}
\end{align}
\end{subequations}
We shall view the first equation as an ODE for $u$ to be solved
in the ``time'' variable $x$. It is slightly more convenient to
use $\tau=x/\epsilon$ as our ``time'' variable.
Rewrite the first equation as a system of first order ODEs:
\begin{subequations}\label{uweqn}
\begin{align}
u_\tau&=w,\\
w_\tau&=-f(u,v).
\end{align}
\end{subequations}
Note that this ODE system possesses an ``energy''.
Multiplying both sides of \eqref{fD} by $w=u_\tau$ and integrating, we obtain
\begin{equation}
w^2=F(u,v,B),
\quad F(u,v,B)=-B+F_0(u,v),\quad F_0(u,v)=-\int_0^u 2f(s,v)ds \label{energy}
\end{equation}
where $B$ is an integration constant.
Consider the $u-w$ phase plane that corresponds to system \eqref{uweqn}.
The solution curves of \eqref{uweqn} coincide with the level curves
of the function $w^2=F_0(u,v)$ (see Fig. \ref{phaseplane}).
Recall that the function $f(u,v)$ is bistable in $u$ for fixed $v$ satisfying
$v_\text{min}<v<v_\text{max}$ (the bistability condition). We shall
be interested only in analyzing cases in which $v$ falls
within this bistable range.
In this case, the function $y=F_0(u,v)$ for fixed $v$
has the form of a
double well potential, whose local minima are at $u=u_+,u_-$
and whose local maximum is at $u=u_m$.
A stationary solution of system \eqref{model}
satisfying equation \eqref{nofluxux} corresponds to a solution trajectory
in the $u-w$ phase plane
that starts and ends at the $u$-axis (or $w=u_\tau=\epsilon u_x=0$).
It is clear that there can only be such a trajectory if $B=F_0(u,v)$
as an equation for $u$ has four distinct solutions
(see Fig. \ref{phaseplane}). Let the two
middle roots be $u_0$ and $u_1$ (we assume $u_0<u_1$). Then,
stationary single front solutions
correspond to the ``half loop'' trajectory
that connects $(u,w)=(u_0,0)$ and $(u_1,0)$.
We see immediately that such stationary single front solutions must be
either monotone increasing or decreasing. In fact, the only
stationary solutions \eqref{model} can have are multiple ``half loop''
trajectories that correspond to periodic multiple front solutions.

For every $(v,B)$ such that $F(u,v,B)=0$ has four solutions in $u$,
we have a corresponding half loop trajectory.
We have, therefore, a two parameter family of half loop trajectories,
and hence of possible stationary single front solutions.
In the following, we shall simply refer to stationary single front
solutions as front solutions.
Let $\mathcal{D}_{vB}$ be the range
of $(v,B)$ values for which $F(u,v,B)=0$ has four solutions.
It is clear that (see Fig. \ref{phaseplane}):
\begin{equation}
\begin{split}
\mathcal{D}_{vB}&=\lbrace(v,B)\in \mathbb{R}^2 |
v_\text{min}<v<v_\text{max}, \;
B_\text{min}(v)<B<B_\text{max}(v)\rbrace\\
B_\text{min}(v)&=\max F_0(u_\pm(v),v),\; B_\text{max}(v)=F_0(u_m(v),v)
\end{split}
\label{vBrange}
\end{equation}
The expression for $B_\text{max}$ denotes the greater of the values
$F_0(u_+(v),v)$ and $F_0(u_-(v),v)$. We thus have a correspondence
between $(v,B)$ values in $\mathcal{D}_{vB}$ and half-loop trajectories.
The set $\mathcal{D}_{vB}$ exhausts all possible half-loop trajectories
but one cannot say in general whether this correspondence is one-to-one.
However, we do expect this to be true if
$f$ is not too pathological.
For reaction terms \eqref{cubic}, \eqref{cubicmod} or \eqref{hilldimless},
it is not difficult to see that this is indeed the case.
We shall henceforth
assume that this correspondence is one-to-one.

The task of finding front solutions has now been reduced
to finding the suitable half-loop trajectories that satisfy
\eqref{nofluxux} and \eqref{uvKinf}.
First, consider \eqref{nofluxux}.
Half-loop solutions automatically satisfy $w=u_\tau=0$ and hence $u_x=0$
at the endpoints, but
this does not necessarily mean that the endpoints correspond to
$x=0$($\tau=0$)
and $x=1$($\tau=1/\epsilon$).
We must thus impose the condition that the domain length is $1$.
Suppose the
front solution has value $u_0$ at $x=0$ and $u_1$ at $x=1$, $u_0<u_1$
(we shall henceforth assume that our front solution
is always monotone increasing, unless noted otherwise).
The domain length condition reduces to:
\begin{equation}
1=\int_0^1 dx=\int_{u_0}^{u_1} \D{x}{u}du=\epsilon
\int_{u_0}^{u_1} \frac{du}{\sqrt{F(u,v,B)}}\equiv \epsilon I(v,B)
\label{lengthAB}
\end{equation}
where we used \eqref{energy}. The above change of variables is
valid because we know that the stationary front solution is
monotone increasing. Note that $u_0$ and $u_1$, being the middle
roots of the equation $F(u,v,B)=0$, are functions of $v$ and $B$.
Hence, the above integral is a function of $v$ and $B$.
Condition \eqref{uvKinf} can, likewise, be written in the following form.
\begin{equation}
K=v+\epsilon \int_{u_0}^{u_1}\frac{udu}{\sqrt{F(u,v,B)}}
\equiv v+\epsilon J(v,B).
\label{massAB}
\end{equation}
We have thus reduced \eqref{Dinftysteady} to
the two integral constraints \eqref{lengthAB} and \eqref{massAB}.
Furthermore, the integral constraints incorporate the fact that
we seek single-front solutions; \eqref{Dinftysteady} is satisfied
by any stationary solution.
Given $\epsilon$ and the total mass $K$, we may solve \eqref{lengthAB}
and \eqref{massAB} for $v$ and $B$,
which in turn uniquely determine the half-loop
trajectory, and hence, the solution $u$.
We note that a similar reduction is possible even when $D$
is finite. We describe this in Appendix \ref{appendix_finiteD}. When $D$ is finite,
however, it seems somewhat difficult to use this reduction to
great effect.

Conditions \eqref{lengthAB} and \eqref{massAB} may be used in place
of \eqref{Dinftysteady} as the basis
for continuing the pinned solution. The use of conditions
\eqref{lengthAB} and \eqref{massAB} has a computational
advantage over the direct use of \eqref{Dinftysteady}
since the former is a much smaller system to solve than the latter.
We do note, however, that the numerical evaluation of the integrals $I(v,B)$
and $J(v,B)$ is not entirely trivial, especially when $B$ is close to
$B_\text{min}$. This is related to the fact that the half-loop trajectories
come very close to heteroclinic or homoclinic orbits on the $u-w$ phase
plane. The techniques used to overcome this difficulty are discussed in
\cite{jilkine_phdthesis}.

A more interesting use of the above conditions is the following.
Since $\epsilon\neq 0$, we may eliminate $\epsilon$ from
\eqref{massAB} and \eqref{lengthAB}. We have:
\begin{equation}\label{QKvB}
Q_K(v,B)\equiv(K-v)I(v,B)-J(v,B)=0.
\end{equation}
If we can find the zero set of $Q_K(v,B)$ where
$(v,B)\in \mathcal{D}_{vB}$,
we will have obtained {\em all} single front
stationary solutions for a fixed value of $K$
(with $v$ in the bistable range)
regardless of whether it arises as a
continuation of the wave-pinned solution.

Any point on this zero set corresponds to a different front solution,
and the value of $\epsilon$ can be recovered by using \eqref{lengthAB}.
Consider the map:
\begin{equation}
\mathcal{M}:(v,B) \longmapsto (M,\epsilon)=(M(v,B),(I(v,B))^{-1})\label{mapM}
\end{equation}
where the function $M(v,B)$ is chosen so that
the map $\mathcal{M}$ defines a homeomorphism on $\mathcal{D}_{vB}$.
Note that the choice of $M$ is far from unique; we shall see that $M(v,B)=v$
works well for \eqref{cubic} and \eqref{cubicmod}.
Half-loop trajectories can then be
parametrized by $(M,\epsilon)$ instead of $(v,B)$.
The zero-set of $Q_K(v,B)$ in $\mathcal{D}_{vB}$ can be mapped by $\mathcal{M}$
in a one-to-one fashion
to yield a bifurcation curve on the $M-\epsilon$ plane.

% INSERTED:
Up to now, the treatment has been fully general.
We now apply this methodology to the case when the reaction term
is given by \eqref{cubic}.
We shall be interested in obtaining the bifurcation diagram when $1<K<3$,
the wave-pinning regime (see \eqref{1K3}).
First, we note that $0<v$ is the bistable range. The domain $\mathcal{D}_{vB}$
is therefore an unbounded set, making it difficult to uniformly
sample points in $\mathcal{D}_{vB}$ to determine the zero set of $Q_K(v,B)$
and hence the bifurcation curves. The following considerations
allow us to restrict our search to a much smaller set.
Given that $u_0$ and $u_1$ are the
two middle roots of equation $F(u,v,B)=0$, we have:
\begin{equation}
0=u_-(v)<u_0<u_m(v)=1<u_1<u_+(v)=1+v.
\end{equation}
Therefore,
\begin{equation}
0<u_0<\int_0^1 u dx<u_1<1+v.
\end{equation}
Using \eqref{uvK},
\begin{equation}
v<v+\int_0^1 udx=K<1+2v.
\end{equation}
Therefore, we may restrict our search of the zero set of $Q_K(v,B)$
to the following range:
\begin{equation}
\frac{K-1}{2}<v<K.\label{vrange}
\end{equation}
We note in passing that a similar argument can be used to
show that {\em any} single front stationary solution to \eqref{Dinfty}
without restriction on $v$ (for $1<K<3$) must in fact satisfy
$v>0$ and hence \eqref{vrange}.

We thus numerically evaluate $Q_K(v,B)$ at sample
points in the range $\mathcal{D}_{vB}^K=\mathcal{D}_{vB}\cap
\lbrace\frac{K-1}{2}<v<K\rbrace$ to find the zero set of $Q_K(v,B)$.
More specifically, we fix $v$ and sample $B$ uniformly within
the admissible range. If there are adjacent sample $B$ points
for which $Q_K(v,B)$ changes sign, a zero is obtained between
these values by bisection. This procedure is repeated for $v$
values uniformly sampled in \eqref{vrange}. Where the zero set
has a complicated structure, sampling is refined to
clarify this structure.
Once the zero-set is obtained, we use the map $\mathcal{M}$ with $M(v,B)=v$
(see \eqref{mapM})
to obtain a bifurcation curve in the $v-\epsilon$ plane.
Computational evidence indicates that $\epsilon=(I(v,B))^{-1}$ is
an increasing function of $B$ for fixed $v$, and thus
$\mathcal{M}:(v,B)\longmapsto (v,\epsilon)$
is a homeomorphism on $\mathcal{D}_{vB}$.

We can explicitly obtain the region
$\mathcal{D}_{v\epsilon}=\mathcal{M}(\mathcal{D}_{vB})$ by
studying the integral $I(v,B)$. Assuming that $I(v,B)>0$ is a decreasing
function of $B$ for fixed $v$, we have only to know the limiting values
of $I(v,B)$ as $B\rightarrow B_\text{min}(v)$ and
$B_\text{max}(v)$ (see \eqref{vBrange}). As
$B\rightarrow B_\text{min}$ for fixed $v$,
the half loop trajectories approach
(half of) a homoclinic orbit or a heteroclinic orbit in the $u-w$ plane
(see Fig. \ref{phaseplane}).
In either case, the total ``time'' it takes for the orbit to
complete the half loop increases as $B\rightarrow B_\text{min}$.
Thus, $I\rightarrow \infty$ as $B\rightarrow B_\text{min}$.
On the other hand, when $B\rightarrow B_\text{max}$, the half-loop
trajectory approaches the neutral center $(u,w)=(u_m,0)=(1,0)$
in the $u-w$ phase plane. We may easily compute:
\begin{equation}
\lim_{B\nearrow B_\text{max}}I(v,B)=\frac{\pi}{\sqrt{v}}.\label{piv}
\end{equation}
Therefore:
\begin{equation}
\mathcal{D}_{v\epsilon}=\lbrace(v,\epsilon)\in \mathbb{R}^2|
0<v, 0<\epsilon <\sqrt{v}/\pi\rbrace.\label{Dveps}
\end{equation}
As one approaches the parabolic edge of $\mathcal{D}_{v\epsilon}$,
the amplitude($=u_1-u_0$) of the front solution tends to $0$ and approaches
the spatially homogeneous steady state $(u,v)=(1,v)$.
In fact, the parabolic edge is the only place where the amplitude
tends to $0$ in $\mathcal{D}_{v\epsilon}$.
Combining this with \eqref{Dveps} with
\eqref{vrange}, we obtain an upper
bound $\epsilon<\sqrt{K}/\pi$ for the existence of
single front solutions. As we shall see, this bound is not sharp.

To study the stability of the stationary solutions corresponding to
points on the zero set,
we must compute $u$ explicitly. Once we know $v$ and $B$, we can find
$u_0$ and $u_1$. We can then numerically solve the initial value problem
\eqref{uweqn} with the initial values $u(0)=u_0$ and $w(0)=0$.
Up to numerical error,
the computed solution should, by design, satisfy
$u(\tau=1/\epsilon)=u(x=1)=u_1$ and
$w(1/\epsilon)=u_\tau(\tau=1/\epsilon)=\epsilon u_x(x=1)=0$.
We can then linearize about $u$ the operator on the right hand side of
\eqref{Dinfty}. By examining the spectrum of (the discretization of)
this linearized operator, we can determine linear stability of the steady
state $u$.

\begin{figure}
\begin{centering}
\includegraphics[width=\textwidth]{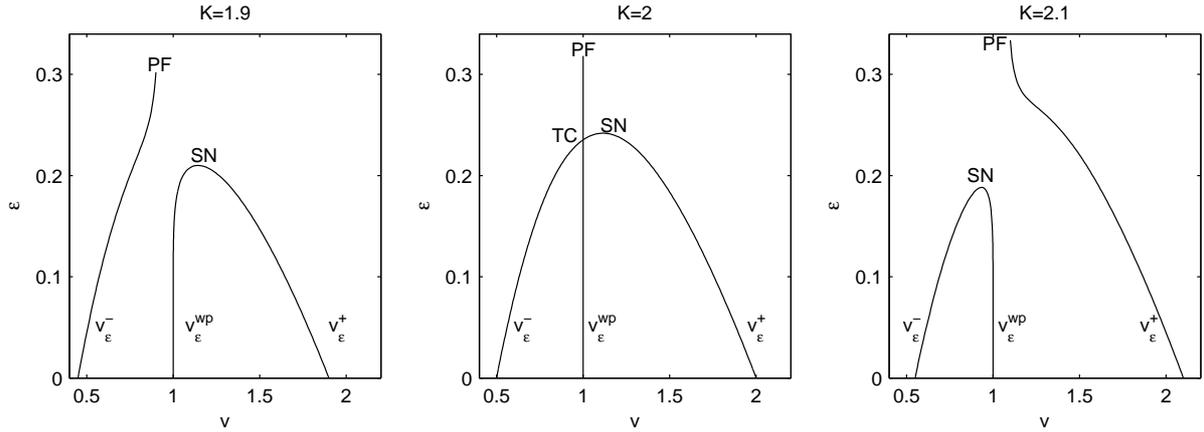}
\end{centering}
\caption{\footnotesize{
Bifurcation diagrams for cubic kinetics \eqref{cubic} 
in the $v-\epsilon$ plane for $K<2$, $K=2$ and $K>2$
(from left to right).
%Three representative bifurcation diagrams for $K<2$, $K=2$ and $K>2$ respectively when \eqref{cubic} is used for the reaction term. The bifurcation curves are plotted in the $v-\epsilon$ plane.
When $K\neq 2$, the middle branch
is stable and the others are unstable, except for a small region 
$2<K<K_p\approx 2.00672$ (details in text and
Fig. \ref{bif2001}).
When $K=2$, middle and minus branches meet at
a transcritical bifurcation (TC,
$(v_\text{tc},\epsilon_{tc})\approx(1,0.23530)$) and exchange stability.
The values $v^-_\epsilon$ and $v^+_\epsilon$ tend to $(K-1)/2$ and $K$,
respectively, as $\epsilon\rightarrow 0$. The slopes of the line of tangency
of the curves $(v^-_\epsilon,\epsilon)$ and $(v^+_\epsilon,\epsilon)$
as $\epsilon\rightarrow 0$ are calculated
in \eqref{vmeps} and \eqref{vpeps}.
The pitchfork bifurcation (PF) occurs at $(v_\text{pf},\epsilon_\text{pf})=(K-1,\sqrt{K-1}/\pi)$. Note that the above bifurcation diagram only shows the
monotone increasing front solution. At the pitchfork bifurcation, this
meets with the monotone decreasing front as well as the spatially
homogeneous state.
SN: saddle-node bifurcation.
}}
\label{bifdiagram}
\end{figure}

The resulting bifurcation curves on the $v-\epsilon$ plane
are given in Fig. \ref{bifdiagram}.
When $K\neq 2$, we found that there was at most one front solution that
corresponds to each value of $v$ (or equivalently, $Q_K(v,B)=0$ had
at most one solution in $B$ for fixed $v$).
We first consider the case $K<2$ (left panel).
For small values of $\epsilon$,
there are three front solutions. In order
of increasing $v$, we denote these solutions
$(u,v)=(u^-_\epsilon(x),v^-_\epsilon), (u^\text{wp}_\epsilon(x), v^\text{wp}_\epsilon)$ and $(u^+_\epsilon(x),v^+_\epsilon)$, which we call 
the minus, middle and plus branches respectively.

The pinned solution corresponds to the middle branch
$(u^\text{wp}_\epsilon(x), v^\text{wp}_\epsilon)$.
The value $v^\text{wp}_\epsilon$ approaches
$1$ as $\epsilon\rightarrow 0$. We know from our asymptotic calculations that
the integral $\int_{u_0}^{u_1}f(u,v)du$ vanishes to leading order
when the wave stalls.
This happens when the three roots $u_{\pm}(v)$ and $u_m(v)$ are equally
spaced, which corresponds to $v=1$.
In the $u-w$ phase plane, $u^\text{wp}_\epsilon$ approaches a heteroclinic
orbit that connects the two saddle points $(u,w)=(0,0)$ and $(u,w)=(2,0)$.

The other two front solutions
are unstable and have a one-dimensional unstable direction.
The values $v^-_\epsilon$ and $v^+_\epsilon$ approach
$v=(K-1)/2$ and $v=K$ respectively as
$\epsilon \rightarrow 0$. Let us consider the plus branch. 
As $\epsilon\rightarrow 0$,
$B$ approaches $B_\text{min}$.
In the $u-w$ phase plane, $u^+_\epsilon$ approaches (half of) a
homoclinic orbit that originates and ends at the saddle point $(u,w)=(0,0)$.
As $u^+_\epsilon$ approaches this homoclinic orbit, the amount of
``time'' that the solution stays close to the saddle point increases,
so that $u^+_\epsilon$ is very close to $0$ for much of the
interval $0<x<1$.
Near $x=1$, there is a sharp transition zone in which
$u^+_\epsilon$
makes a steep increase to $u_1$. This transition zone becomes increasingly
narrow as $\epsilon\rightarrow 0$.
We may say that the solution
$(u^+_\epsilon(x),v^+_\epsilon)$ approaches the stable homogeneous
steady state $(u,v)=(0,K)$ as $\epsilon\rightarrow 0$.
The convergence
of $u^+_\epsilon(x)$ to $0$ is only uniform outside of an arbitrarily
small neighborhood of $x=1$.
We can use the above phase plane information together
with \eqref{massAB} to find the following
approximate expression for $v^+_\epsilon$ when $\epsilon$ is small:
\begin{equation}
\begin{split}
v^+_\epsilon&=K-\alpha_+\epsilon +o(\epsilon),\\
\alpha_+&=\lim_{B\searrow 0}J(K,B)
=2\sqrt{2}
\ln\paren{\frac{\sqrt{\beta_+}+\sqrt{\beta_-}}{\sqrt{\beta_+-\beta_-}}},\\
\beta_{\pm}&=\frac{1}{3}\paren{2(2+K)\pm\sqrt{2(2K+1)(K-1)}}.
\end{split}\label{vpeps}
\end{equation}
where $o(\epsilon)$ is the usual Landau symbol denoting an expression
that tends to $0$ as $\epsilon\rightarrow 0$. Note that the
expression inside the square root in $\beta_\pm$ is positive since $K>1$.
The validity of this expression is supported by computational results.

The situation for the minus branch is similar.
As $\epsilon\rightarrow 0$, $v^-_\epsilon \rightarrow (K-1)/2$.
In the $u-w$ phase plane, $u^-_\epsilon$ approaches half of
the homoclinic orbit originating from the saddle point $(u,w)=((K+1)/2,0)$.
The value of $u^-_\epsilon$ is close to $(K+1)/2$ for
most of $0<x<1$ except for
a small neighborhood around $x=0$.
The function $u^-_\epsilon$
converges uniformly to $(K+1)/2$ on any set outside an arbitrarily
small neighborhood around $x=0$.
Similarly to \eqref{vpeps}, we can obtain the following expression
for $v^-_\epsilon$:
\begin{equation}
\begin{split}
v^-_\epsilon&=\frac{K-1}{2}+\alpha_-\epsilon +o(\epsilon),\\
\alpha_-&=\sqrt{2}
\ln\paren{\frac{\sqrt{\gamma_+}+\sqrt{\gamma_-}}{\sqrt{\gamma_+-\gamma_-}}},\;
\gamma_{\pm}=\frac{1}{3}\paren{2K\pm\sqrt{\frac{1}{2}(3+K)(3-K)}}.
\end{split}\label{vmeps}
\end{equation}
Note that the expression inside the square root in $\gamma_\pm$ is
positive since $K<3$.
The validity of the above expression is supported by numerical results.

As $\epsilon$ is increased, there is a value $\epsilon=\epsilon^+_\text{sn}$
at which the middle and plus branches meet in a saddle-node bifurcation.
At this point, the disappearance of the stable front solution 
(the middle branch) is ``abrupt''
in the sense that the amplitude of $u$ is non-zero as the bifurcation
is approached. This can be seen from the fact that
this saddle-node bifurcation occurs in the interior of
$\mathcal{D}_{v\epsilon}$ (see discussion after \eqref{Dveps}).

The minus branch can be continued
until it merges with a spatially homogeneous unstable steady state.
This happens at the parabolic edge of $\mathcal{D}_{v\epsilon}$
(see \eqref{Dveps}).
Here, there is a pitchfork bifurcation
at which the homogeneous steady state
gives rise to {\em two} unstable front solutions, one that is monotone
increasing and the other monotone decreasing.
Note that, in Fig. \ref{bifdiagram}, only the bifurcation diagram
of the monotone increasing front solution is plotted.
There is an identical bifurcation diagram for the monotone decreasing
front, and these two solutions meet with a spatially homogeneous
steady state at a pitchfork bifurcation.
This homogeneous steady state
corresponds to $(u,v)=(1,K-1)$.
Using \eqref{Dveps}, the corresponding $\epsilon$ value
$\epsilon_\text{pf}$ is equal to:
\begin{equation}
\epsilon_\text{pf}=\frac{\sqrt{v}}{\pi}=\frac{\sqrt{K-1}}{\pi}.\label{epspf}
\end{equation}
Linearize the operator on the right hand side
of \eqref{Dinfty} around this homogeneous steady state, and call
this operator $\mathcal{L}$. We can also obtain the above
value by considering the spectrum of $\mathcal{L}$.
At $\epsilon=\epsilon_\text{pf}$, one of the eigenvalues of $\mathcal{L}$
corresponding to the wave number $k=\pi$
becomes positive (see Section \ref{stability},
in particular, equation \eqref{epsbound} in the $D\rightarrow \infty$ limit).
For $1<K<3$, expression \eqref{epspf} gave the
least upper bound of the range of $\epsilon$ for which
(not necessarily stable) single front solutions exist.

We now turn to the case $K=2$. Let us first take a look at
\eqref{QKvB}. If $v=1$, we have:
\begin{equation}
Q_2(1,B)=\int_{u_0}^{u_1}\frac{1-u}{\sqrt{F(u,1,B)}}du=0.
\end{equation}
The function $F(u,1,B)$ is symmetric about $u=1$ and thus the
same is true for $u_0$ and $u_1$ (i.e. $(u_0+u_1)/2=1$).
The above integral is therefore equal to $0$ whenever it is
well-defined. Therefore, all points such that $v=1$
in $\mathcal{D}_{v\epsilon}$
(i.e., $(v,\epsilon)=(1,\epsilon), 0<\epsilon<1/\pi$)
are part of the bifurcation curve for $K=2$.
For $\epsilon$ small, this $v=1$ branch of solutions
corresponds to the pinned front, which is thus stable 
for small $\epsilon$. We shall denote this branch by 
$(u^\text{wp}_\epsilon,v^\text{wp}_\epsilon)$ and call this the middle 
branch.
As $\epsilon\nearrow 1/\pi$,
we expect the middle branch to merge with an
unstable homogeneous steady state at a pitchfork bifurcation, just
like $u^-_\epsilon$ for $K<2$.
An unstable solution cannot give rise
to two stable solutions in a pitchfork bifurcation.
We must conclude that the middle branch is
unstable when $\epsilon$ is close to $1/\pi$.
This suggests that there must be an intermediate $\epsilon$
value between $0$ and $1/\pi$ at which the middle branch loses
stability. This is indeed the case.

Just as in the $K<2$ case, there are three single front solutions
in the $K=2$ case for small $\epsilon$. We shall refer to them
in the same way as in the $K<2$ case. As we saw, $v^\text{wp}_\epsilon$
is always equal to $1$. As $\epsilon$ is increased, 
the minus and middle branches
meet in a transcritical bifurcation at
$\epsilon=\epsilon_\text{tc}\approx 0.2353$. Above $\epsilon_\text{tc}$,
the minus branch becomes stable and the middle branch loses stability.
At $\epsilon=\epsilon^+_\text{sn}\approx 0.2419$, the 
minus and plus branches meet in a saddle-node bifurcation.

The $K>2$ case is similar to the $K<2$ case except for some fine details.
When $\epsilon$ is small, we have three front solutions which we name
in the same fashion as in the $K\leq 2$ cases. The middle branch 
merges with the minus branch
at $\epsilon=\epsilon^-_\text{sn}$ in a saddle-node bifurcation.
The branch plus branch merges with the spatially
homogeneous solution at $\epsilon_\text{pf}=\sqrt{K-1}/\pi$ (see \eqref{epspf})
in a pitchfork bifurcation.

\begin{SCfigure}
\begin{centering}
\includegraphics[width=0.7\textwidth]{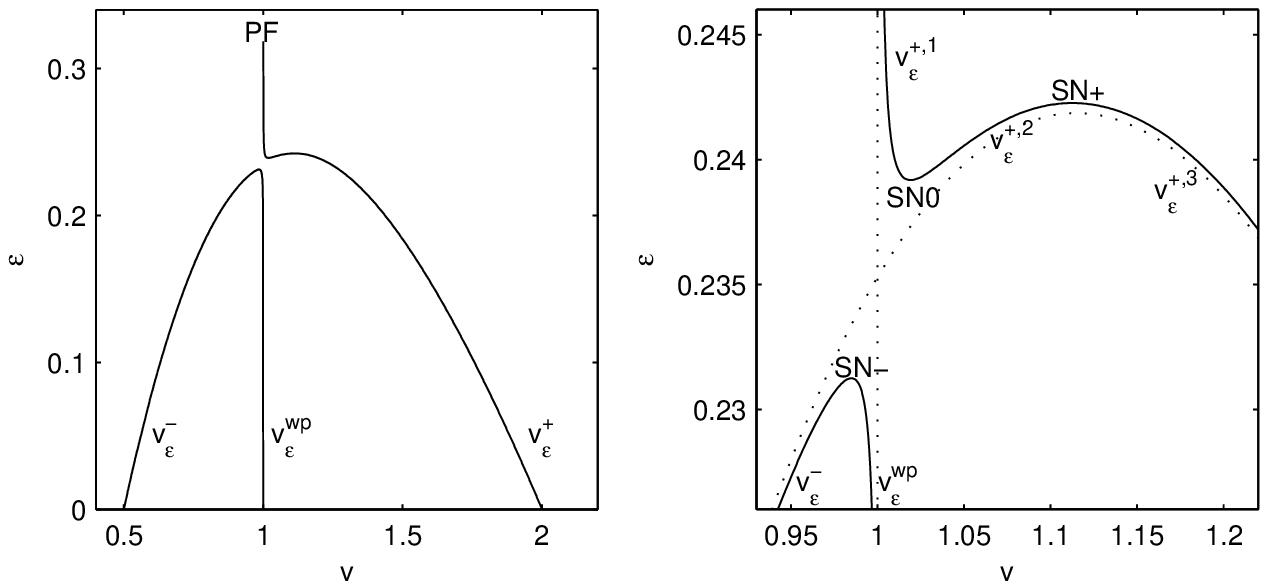}
\end{centering}
\caption{\footnotesize{Bifurcation diagram for cubic kinetics \eqref{cubic} 
for $2<K<K_p\approx 2.00672$
($K=2.001$ in this example) plotted in the $v-\epsilon$ plane.  
The full diagram is on the left panel, a part of
which is magnified on the right panel. The $(+,1)$ and $(+,2)$ 
branches (labeled $v^{+,1}_\epsilon$ and $v^{+,2}_\epsilon$ 
respectively) come together
at the saddle-node bifurcation denoted $SN0$ and the branches 
for $(+,2)$ and $(+,3)$ branches 
(labeled $v^{+,2}_\epsilon$ and $v^{+,3}_\epsilon$ 
respectively)
come together
at $SN+$. The $(+,2)$ branch is stable.
At $SN-$, middle and minus branches come together.
The dotted lines are the bifurcation curves at $K=2$.
}}
\label{bif2001}
\end{SCfigure}

An interesting detail in the $K>2$ case is that there is a small
window $2<K<K_p\approx 2.00672$ for which the 
plus branch has a stable portion (see Fig. \ref{bif2001}). The existence of
such a portion is implied by the structure of the bifurcation diagram
at $K=2$. The saddle-node bifurcation at $\epsilon^+_\text{sn}$ should
persist beyond $K=2$ since saddle-node bifurcations are robust under
perturbations. On the other hand,
when a transcritical bifurcation is perturbed, it will generally give
rise to zero or {\em two} saddle node bifurcations 
(see \cite{kuznetsov2004elements} for example). When $K=2$ is
perturbed to $K<2$, the transcritical bifurcation does not give
rise to any saddle node bifurcations. If perturbed to $K>2$, it gives
rise to two saddle node bifurcations,
one of which occurs at $\epsilon=\epsilon^-_\text{sn}$.
We shall name the other $\epsilon$ value $\epsilon=\epsilon^0_\text{sn}$.
Both bifurcation points corresponding to $\epsilon=\epsilon^0_\text{sn}$
and $\epsilon^+_\text{sn}$ lie on the
$(u^+_\epsilon,v^+_\epsilon)$ branch.

For $2<K<K_p$, there are three single front solutions over the range
$\epsilon^0_\text{sn}<\epsilon<\epsilon^+_\text{sn}$, which we 
refer to as the $(+,1),(+,2)$ and $(+,3)$ branches respectively
in order of increasing $v$.
The $(+,1)$ and $(+,2)$ branches meet in a saddle-node
bifurcation at $\epsilon^0_\text{sn}$ and
$(+,2)$ and $(+,3)$ branches meet at $\epsilon^+_\text{sn}$.
The $(+,1)$ and $(+,3)$ branches are unstable whereas
$(+,2)$ branch is stable. For $2<K<K_p$, therefore, there is a
small window of $\epsilon$ values for which there is a stable front solution
that cannot be reached by continuing the pinned front solution.
For $K\geq K_p$, the plus branch does not have a stable portion.
At $K=K_p$, the saddle-node bifurcation points merge and disappear.

\begin{SCfigure}
\begin{centering}
\includegraphics[width=0.7\textwidth]{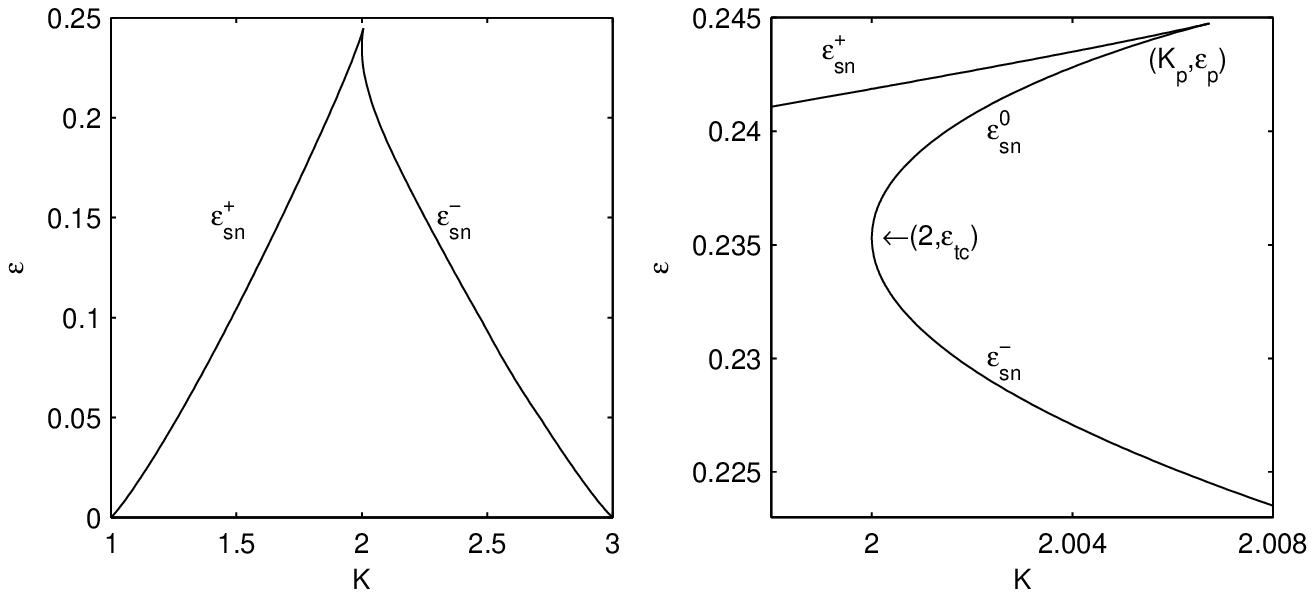}
\end{centering}
\caption{\footnotesize{
Two parameter bifurcation plots for cubic kinetics \eqref{cubic}. 
A stable front solution exists for parameter values
in the $K-\epsilon$ parameter region bounded by the curve and the $K$-axis
(left panel). On the right panel, the ``tip'' of the curve is magnified.
At $(K_p,\epsilon_p)\approx(2.00672,0.24474)$ the values $\epsilon^+_\text{sn}$
and $\epsilon^0_\text{sn}$ come together. At $(2,\epsilon_\text{tc}), \epsilon_\text{tc}\approx 0.23250$, $\epsilon^-_\text{sn}$ and $\epsilon^0_\text{sn}$
come together at the transcritical bifurcation point.
}}
\label{epsDinfty}
\end{SCfigure}

In Fig. \ref{epsDinfty}, we show the $(K,\epsilon)$ parameter region in which
there is a stable single front solution.
This should be seen as a refinement of the $\epsilon_c$
plot in Fig. \ref{epsDfinite} that we obtained for finite $D$.
The region is peaked at approximately
$K=2$, but with some fine structure coming from the small window of
front solutions that exist for $2<K<K_p$.
The peaked geometry of this region comes from the
fact that the saddle-node bifurcations at which the pinned solution
loses stability are different for $K>2$ and $K<2$. At $K=2$, we have
a transcritical bifurcation that separates these two regimes.

It is not clear how much of the insights we obtained for
$D\rightarrow \infty$ can be carried over to the finite $D$ case
or to the reaction term \eqref{hill01}.
It seems plausible, however, that much of what we learned
does indeed carry over.
For example, we expect that there is a
$K$ value (that depends on $D$) at which the pinned solution undergoes
a transcritical bifurcation (rather than a saddle-node bifurcation) as
$\epsilon$ is increased.
The peaked appearance of the $\epsilon_c$ plot
of Fig. \ref{epsDfinite} serves as circumstantial evidence for this claim.

\subsection{Other Possible Bifurcation Structures}\label{otherbif}

We now have a clear picture of the bifurcation structure for
 reaction term \eqref{cubic}, especially when $D\rightarrow \infty$.
Given the broad similarity of the $\epsilon_c$ plots for \eqref{hill01} and \eqref{cubic}
(see Figure \ref{epsDfinite}), 
it is natural to expect \eqref{hill01} to also have a bifurcation structure 
with features similar to \eqref{cubic}.
This raises the question of how general our findings are.
For other reaction terms that support wave-pinning, there is
no reason to expect the full bifurcation structure to be similar.
In particular, we can raise the following question. Except at $K=2$,
the pinned front was seen to undergo a saddle-node bifurcation
in the case of \eqref{cubic}, $D\rightarrow \infty$.
This bifurcation was ``abrupt'' in the sense that the front amplitude
tends to a non-zero value as the bifurcation point is approached.
Is the saddle-node bifurcation the only generic way in which the
pinned front is lost?
In particular, is it generically the case that the bifurcation is ``abrupt''?
The answer to both questions turn out to be negative.
We shall demonstrate this with a description of the
bifurcation structure for the reaction term \eqref{cubicmod}.
We shall see that the pinned front can arise
generically via a pitchfork bifurcation from a spatially homogeneous state.
The exposition will be kept brief since much of the analysis proceeds
along lines similar to that of the previous section.
We shall only discuss the $D\rightarrow \infty$ case.
The case of $D$ finite is expected to be similar. We note in
particular that computational examples can be produced in which
such bifurcations occur for finite $D$.

As we saw in Section
\ref{stability}, an interesting feature of the reaction term
\eqref{cubicmod}
is that the middle homogeneous steady state $(u_m,v)$ can be stable.
A similar conclusion is true in the $D\rightarrow \infty$ case.
This happens when $a>1$ in \eqref{cubicmod}. We shall concentrate
on this case. When $a<1$, the full bifurcation diagram turns out
to be quite similar to that of \eqref{cubic} (the generic bifurcation
is the saddle-node).
We shall not discuss this case here.

As can be easily checked, $-\infty<v<\infty$ is the bistable
range, and $-1<K<1$ is the range for which wave-pinning can
occur. We focus on these values of $K$.

Let us first study the spatially homogeneous steady states of
the system for fixed $K$. Given $v$, $u$ must be either
$u=u_+,u_-$ or $u_m$. There is one spatially homogeneous steady
state each for $u_-$ and $u_+$:
$(u_-,v)=(-1,K+1)$ and $(u_+,v)=(1,K-1)$.
Let us consider the spatially homogeneous steady states that
correspond to $u=u_m$. Given that the total mass must equal $K$,
we have the following equality:
\begin{equation}
v+u_m(v)=v-\frac{av}{\sqrt{1+(av)^2}}=K.\label{vequm}
\end{equation}
It turns out that this equation can have three solutions in $v$ for
fixed $K$ if $a>1$ and $K$ satisfies:
\begin{equation}
-K_q<K<K_q,\; K_q=\frac{1}{a}(a^{2/3}-1)^{3/2}.\label{Kq}
\end{equation}
It is clear that $K_q$ is always smaller than $1$.
Let us call these three solutions $v_m^-<v_m^0<v_m^+$.
We may adapt the calculations of Section \ref{stability}
to the $D\rightarrow\infty$ case.
It can be checked that $\tau_0=f_u-f_v<0$ at
$(u_m^0,v^0_m)\equiv(u_m(v_m^0),v_m^0)$, and therefore,
that this is a stable steady state so long as:
\begin{equation}
\epsilon>\frac{\sqrt{f_u(u_m^0,v_m^0)}}{\pi}\equiv\epsilon^0_{\text{pf}}
\label{eps0pf}
\end{equation}
Note that the above expression can be obtained by taking
$D\rightarrow \infty$ in \eqref{epsbound}. We name the right hand side
$\epsilon^0_\text{pf}$ in anticipation of our results to be discussed below.
The other two homogeneous states are always unstable.
When $\abs{K}>K_q$ \eqref{vequm} has only one solution.

\begin{figure}
\begin{centering}
\includegraphics[width=\textwidth]{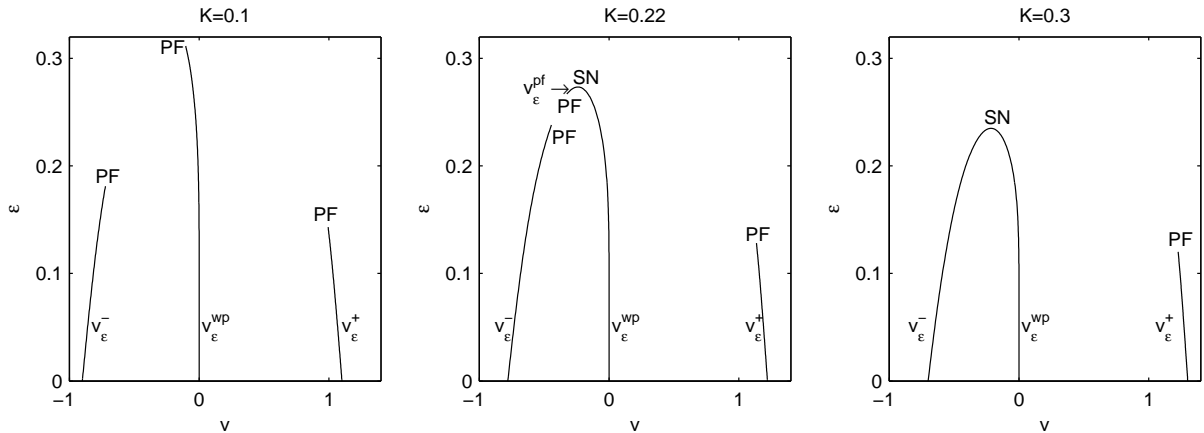}
\end{centering}
\caption{\footnotesize{%In the first three figures 
Bifurcation diagrams for the reaction term 
\eqref{cubicmod} in the $v-\epsilon$ with $a=2$
%The bifurcation curves are plotted in the $v-\epsilon$ plane. 
and indicated values of $K$. Left: $K=0.1<K_r\approx 0.19498$,
%The left panel is for $K=0.1<K_r\approx 0.19498$,
%the middle panel 
Middle: $K_r<K=0.22<K_q\approx 0.22510$,
$v^\text{pf}_\epsilon$
is represented by the small portion of the curve between the pitchfork
bifurcation (PF) and the saddle-node bifurcation (SN). 
Right: $K=0.3>K_q$.
%Bottom right: {\bf FILL THIS IN OR DELETE THAT PANEL.}
%and the right panel $K=0.3>K_q$. In the middle panel, $v^\text{pf}_\epsilon$is represented by the small portion of the curve between the pitchfork bifurcation (PF) and the saddle-node bifurcation (SN). %% In the final 
%% panel, we plot $K_r$ and $K_q$ as a function of $a$. The $a$-axis is 
%% scaled logarithmically. Note how $K_r\rightarrow 1$
%% as $a\rightarrow \infty$. 
}}
\label{bifmod}
\end{figure}

The bifurcation diagram in the $D\rightarrow \infty$ limit can
be obtained in a procedure similar to the treatment of \eqref{cubic}
in the previous section.
The possible bifurcation diagrams in the $v-\epsilon$ plane
are given in Fig. \ref{bifmod}, where we have taken $a=2$ in \eqref{cubicmod}.
Only the case $K\geq 0$ is shown. Given the symmetry of the
reaction term \eqref{cubicmod}, the bifurcation diagram for
$-K$ can be obtained by flipping the bifurcation diagram for $K$
about the $\epsilon$ axis.

For all values of $-1<K<1$, there are three single front solutions
 when $\epsilon$ is sufficiently small. Just as in the previous section, 
we shall denote them by 
$(u^-_\epsilon,v^-_\epsilon),
(u^\text{wp}_\epsilon,v^\text{wp}_\epsilon), (u^+_\epsilon,v^+_\epsilon)$
and refer to them as the minus, middle and plus branches.
There is a constant $0<K_r<K_q$ (that depends on $a$) such that,
when $0\leq K\leq K_r$ the middle branch 
merges with the stable
homogeneous solution $(u^0_m,v^0_m)$ in a pitchfork bifurcation.
This happens at $\epsilon=\epsilon^0_\text{pf}$ whose analytical
expression was given in \eqref{eps0pf}.
The pinned front solution is stable up to
this pitchfork bifurcation. Note that this is only possible
since $(u^0_m,v^0_m)$ is a stable steady state for 
$\epsilon>\epsilon^0_\text{pf}$.
The minus and plus branches are unstable and
merge in pitchfork
bifurcations, respectively, with the unstable homogeneous states
$(u^{\pm}_m,v^{\pm}_m)$.

When $K_r<K<K_q$, the situation for the plus and minus branches 
does not change.
However, the middle branch now loses stability
in a saddle-node bifurcation
with the solution branch $(u^\text{pf}_\epsilon,v^\text{pf}_\epsilon)$
that arises from a pitchfork
bifurcation from the homogeneous state $(u^0_m,v^0_m)$.
This branch is unstable.
The difference between $0\leq K\leq K_r$ and $K_r<K<K_q$ is whether the
pitchfork bifurcation at $(u^0_m,v^0_m)$ is subcritical or
supercritical (see Fig. \ref{bifmod}). In fact, we  
encountered a similar bifurcation for \eqref{hill01}
when $D=1$ and $K=2.9$ (see 
Fig. \ref{samplebif} (b) and (c)).

For $K>K_q$, the middle branch loses stability in a saddle-node
bifurcation with the minus branch. The plus branch
merges with the unstable homogeneous solution $(u^+_m,v^+_m)$.
The case $K=K_q$ is highly degenerate and atypical, and we thus omit
the details here.

Assuming that the above bifurcation picture is valid for all values of $a>1$
(an observation supported by computational evidence), 
we can compute $K_r$ as the value of $K$ at which the pitchfork bifurcation at 
$(u^0_m,v^0_m)$ changes from being subcritical to supercritical.
We can then obtain an explicit analytical expression for $K_r$, whose 
derivation we defer to Appendix \ref{appendix_Kr}.
We note that values of $K_r$ obtained by bifurcation computations
match perfectly with the analytical expression we now state. 
Consider the equation:
\begin{equation}
\frac{5}{3}+\frac{a}{s^3-a}+\frac{3}{8(s^2-1)}=0.\label{seq}
\end{equation}
It can be shown that there is just one solution to the above
equation in the range $1<s<\sqrt[3]{a}$. Take this root
and let:
\begin{equation}
v_r=-\frac{\sqrt{s^2-1}}{a},\; K_r=v_r-\frac{av_r}{s}.\label{Kreq}
\end{equation}
The value $v_r$ is the value of $v_m^0$ when $K=K_r$.
We thus have an expression for $K_r$ as a function of $a$.
We can see that $K_r\rightarrow 1$ as $a\rightarrow \infty$.
Letting $\sigma=s/\sqrt[3]{a}$, we can rewrite \eqref{seq} as:
\begin{equation}
\frac{31\sigma^3-7}{\sigma^2(40\sigma^3-16)}=a^{2/3}.
\end{equation}
We see that $\sigma \rightarrow \sqrt[3]{16/40}=\sqrt[3]{2/5}$ as 
$a\rightarrow \infty$. Using \eqref{Kreq},
\begin{equation}
\lim_{a\rightarrow \infty} K_r
=\lim_{a\rightarrow \infty}\paren{-\frac{\sqrt{s^2-1}}{a}+\frac{\sqrt{s^2-1}}{s}}
=\lim_{a\rightarrow \infty}\paren{-\frac{\sqrt{a^{2/3}\sigma^2-1}}{a}
+\sqrt{\frac{a^{2/3}\sigma^2-1}{a^{2/3}\sigma^2}}}=1.
\end{equation}
In other words, the range of $K$
over which the pinned solution merges with a stable homogeneous
solution increases with $a$, covering the
entire wave-pinning regime ($-1<K<1$) as $a\rightarrow \infty$.
%% In Figure \ref{bifmod}, we plot $K_r$ and $K_q$
%% as a function of $a$.

For reaction term \eqref{cubic}, the only generic bifurcation
through which the pinned solution is lost was of saddle-node type.
Reaction term \eqref{cubicmod} is an example in which the
pinned solution can be generically lost by merging with a
stable spatially homogeneous state. As $a\rightarrow\infty$,
this is the case for most values of $K$ in the wave-pinning regime.
Although these examples give us interesting insight into the possible
bifurcation structure of \eqref{model}, it is difficult to draw
conclusions that may be applicable to arbitrary reaction terms.
Our study in the present section suggests a general connection between
the stability of homogeneous states of type $(u_m(v),v)$
and the type of bifurcation at which the pinned solution
is lost.

\section{Discussion}
Previously we have studied the reaction-diffusion model \eqref{OurModel}  with kinetics \eqref{f Hill function},
motivated by an investigation of the redistribution of polarity
proteins (Rho family GTPases) in eukaryotic cells.
These switch-like proteins interconvert between an active and an inactive
form and diffuse across the cell.
The appearance of a small parameter, $\epsilon$ in this problem stems from
the membrane confinement of one of the species (the active form),
which tends to reduce its rate of diffusion by orders of magnitude
relative to the other form. The inactive form diffuses rapidly, i.e. $D = O(1)$.
Conservation of total amount of protein ($K_{total}$, and in dimensionless form $K$)
stems from the fact that there is no net production nor loss of total
protein on the timescale of interest.

In previous work on this biological problem, we had postulated bistability based on positive feedback
between the presence of the active form and its own activation. This led us to find
a phenomenon of wave-pinning, which could account for polarization
in response to large enough stimuli \cite{mori_bj08}. We found that
the phenomenon depends on the ratio between the two diffusion coefficients being small enough. 
Many other proposed models for cell polarization are based on diffusion-driven, Turing-type instabilities
\cite{subramanian_jtb_04,narang_jtb_05,Otsuji-plos07}, 
in which a state that is stable in the absence of diffusion is destabilized 
in its presence, a mechanism fundamentally different from the wave-pinning 
mechanism considered here.
Our main motivation has been to understand this phenomenon from a mathematical point of view.

We first analyzed wave-pinning exploiting the smallness of $\epsilon$
using matched asymptotic analysis. We identified three key
properties the reaction term must satisfy (bistability, homogeneous stability and the 
velocity sign conditions, see Section \ref{ModelFormul}) 
in order for the system to exhibit wave-pinning. 
Both \eqref{f Hill function}, as well as the simpler \eqref{cubic} satisfies
these properties and thus supports wave-pinning. 
The analysis allowed us to determine the range of $K$ values for which 
wave-pinning is possible. Furthermore, we were able to 
reduce the RD system to a simple differential algebraic 
system for the front position, whose explicit form could be calculated in the case of
\eqref{cubic} thanks to its algebraic simplicity. This reduction gives an excellent 
approximation of the original system as $\epsilon$ is made small (Fig. \ref{Fig:Error_phi_0}). 
We briefly discussed the long-time behavior of our system as well as its higher 
dimensional generalizations. We argued that the long-time behavior is analogous 
to that of the mass-constrained Allen-Cahn model, whose properties have 
been well-characterized 
\cite{rubinstein_imajam_92,Ward_sjam_96,
reyna1995metastable,sun2000dynamics}.

As $\epsilon$ is increased, the matched asymptotic calculations are no longer valid, 
and the pinned front is eventually lost. 
This led us to examine the bifurcation structure of the system.  
For finite $D$, we did so using pseudoarclength continuation on the
full PDE system (Figs.~\ref{samplebif}-\ref{epsDfinite}). Reaction terms
\eqref{hill01} and \eqref{cubic} revealed a similar bifurcation structure.
We found that the pinned front was always lost in a saddle-node (fold) bifurcation, 
and delineated the parameter region in the $K-\epsilon$ plane 
for which wave-pinning was possible (Fig. \ref{epsDfinite}). 
We obtained a complete bifurcation picture for single front solutions 
in the limit $D\to \infty$ for the reaction term \eqref{cubic},
using a method related to the ``time map'' technique \cite{smoller_1981, grindrod_book}.
We found that there is a transcritical bifurcation for a
particular value of $K(=2)$ (Fig. \ref{bifdiagram}). This value acts as a ``watershed'' 
explaining the cusp-like form seen in Fig. \ref{epsDfinite}.
Other bifurcation pictures are possible.
In the case of \eqref{cubicmod}, as shown in Fig.~\ref{bifmod}, 
the pinned front solution can be lost through a pitchfork bifurcation. 
The possibility of such a bifurcation depends 
on the stability of the ``middle'' homogeneous steady states.
It seems to be difficult to give a 
general account of the bifurcation structure for wave-pinning systems.
We hope the two scenarios we identified are representative of what 
can be expected.

The simplicity of our model and the universality of
reaction-diffusion systems in biology, chemistry, and physical settings suggests
that such wave-pinning phenomena may be quite ubiquitous 
\cite{meerson_96, sepulchre_chaos_00,sneyd_1993,wylie_pre_06}.
In this paper, our motivation stems from cell polarization and the biochemistry 
of Rho proteins, and the variables $u$ and $v$ correspond to active and
inactive forms of one Rho protein.
More detailed models for the dynamics of these proteins, with
mutual interactions and effects on the actin cytoskeleton
\cite{jilkine_bmb_07,maree_bmb_06,dawes_bj_07} show similar
wave-pinning phenomena, but their complexity makes
a full mathematical analysis much harder.

We conclude with a discussion of possible biological implications.
The small parameter $\epsilon$ exploited in our analysis
depends on several biological parameters including rates of 
diffusion $D_u$, reaction $\eta$, and domain size $L$.
The necessary condition $\epsilon \ll 1, D\approx O(1)$
is satisfied by virtue of the large difference in diffusion of
the  membrane-bound active Rho protein and inactive form that
diffuses freely in the cytosol.
Normally, these rates of diffusion differ by 100-fold.
Assuming a typical cell diameter of 10 $\mu$m, reaction timescale 
$\eta=1\,$s$^{-1}$, and diffusion coefficients 
$D_u=0.1 \, \mu$m$^2$s$^{-1}$ and $D_v=10 \, \mu$m$^2$s$^{-1}$, 
the dimensionless constants are $\epsilon \approx 0.03$ and $D\approx 0.1$.
This is within the wave-pinning regime for the Hill function kinetics \eqref{f Hill function}.
However, increasing the diffusion coefficient of the active form tenfold
to $D_u=1 \, \mu$m$^2$s$^{-1}$, or slowing down the rate of interconversion $\eta$ to $0.1\,$s$^{-1}$,
or decreasing the cell size to $L \approx 3 \, \mu$m leads to 
$\epsilon \approx 0.1$ and $D\approx 1$, putting the 
Hill function kinetics system into the bifurcation regime
where wave-pinning and hence polarization would be lost. 

Such predictions are experimentally testable. Cell fragments
capable of polarization \cite{Verkhovsky_cb_99} could be made
successively smaller to test the effect of domain size. Manipulating
the amount of Rho protein could test the predicted effect on polarization.
(Some confirmation of the prediction is observed with Cdc42 manipulation in yeast, 
where the frequency of spontaneous polarization 
is inversely dependent on the amount of Cdc42  \cite{Altschuler_08}.)
Replacing a cytosolic protein by a fusion protein with lower mobility 
has been experimentally done in budding yeast \cite{Howell2009}. 
Our results show that reducing $D$ for a fixed $\epsilon$ may lead to loss of polarity,
as the border between wave-pinning and homogeneous regimes is shifted 
(e.g. see Fig.~\ref{epsDfinite}, second panel).
Furthermore, Rho protein cycling between membrane and cytosol
is affected by proteins called GDIs. We have previously shown that the time spent in the
cytosol vs membrane affects the effective diffusion coefficient of the
inactive form $D_v$ \cite{jilkine_bmb_07,maree_bmb_06}, which thus affects the dimensionless parameter $D$.
This suggest that regulation of the GDIs is yet another possible mechanism for
regulating polarity \cite{dernardirossian_mbc_06}).  
Experiments in budding yeast show that knock down (i.e., replacement with a non-functional version) 
of GDI coupled with treatment to disable a second redundant Cdc42 membrane recycling pathway 
leads to rapid dissipation of polarity \cite{Slaughter2009}. This supports our predictions.

\newpage

\section{Appendix}

\subsection{Integral Reduction at Finite $D$}\label{appendix_finiteD}

We perform an integral reduction of \eqref{steadystate} for finite $D$,
similarly to the treatment of the $D\rightarrow\infty$ case in
Section \ref{secDinfty}
We view \eqref{steadystate} as an ODE with $x$ as the ``time'' variable.
First, add (\ref{steadystate}a,b) to obtain
\begin{equation}
\epsilon^2 u_{xx}+Dv_{xx}=0.
\end{equation}
Integrate this equation twice and use the no-flux boundary conditions
to obtain
\begin{equation}\label{Aeqn}
\frac{\epsilon^2}{D} u + v=A,
\end{equation}
where $A$ is an integration constant.
Solving the above for $v$ and substituting this into \eqref{u_steady_state},
we reduce the system to a single equation for $u$:
\begin{equation}\label{fD}
0=\epsilon^2 {u}_{xx}+f_D(u,A),
\end{equation}
where $f_D(u,A)=f(u,A-\epsilon^2u/D)$.
The rest follows along exactly the same lines as in Section \ref{secDinfty}.

We note two differences.
Recall that the function $f(u,v)$ is bistable in $u$ for fixed $v$ satisfying
$v_\text{min}<v<v_\text{max}$.
We can see from \eqref{Aeqn} that if
\begin{equation}
v_\text{min}<A<v_\text{max}
\end{equation}
then $f_D(u,A)$ is bistable in $u$ (for $u$ in a finite range)
for $\epsilon$ small enough, assuming that $f(u,v)$ is a smooth function of $v$. In this case, the function:
\begin{equation}
F_D(u,A,B)=-B-\int_0^u f_D(s,A)ds
\end{equation}
will have the form of a
double well potential, whose local minima correspond to the stable zeros
of the bistable function $f_D(u,A)$.
This restriction on the size of $\epsilon$ was absent in the
$D\rightarrow\infty$ case. This can also be seen by formally taking
the limit as $D\rightarrow \infty$ in \eqref{Aeqn}, which yields $A=v$.

The integral conditions \eqref{lengthAB} and \eqref{massAB}, in the
finite $D$ case have the form:
\begin{align}
1&=\epsilon\int_{u_0}^{u_1} \frac{du}{\sqrt{F_D(u,A,B)}},\\
K&=A+
\epsilon \paren{1-\frac{\epsilon^2}{D}}\int_{u_0}^{u_1}\frac{udu}{\sqrt{F_D(u,A,B)}},
\end{align}
where $u_0<u_1$ are the two middle roots of the equation $F_D(u,A,B)=0$.
It is easy to see that these conditions reduce to \eqref{lengthAB} and \eqref{massAB}
in the $D\rightarrow\infty$ limit.
One difficulty here is that it is not possible to eliminate $\epsilon$ to obtain
a relation analogous to \eqref{QKvB}, since $F_D(u,A,B)$ has an $\epsilon$ dependence.
It is nonetheless possible to use the above as a basis for a continuation algorithm,
and we have seen that the results using these relations match with those obtained
using a direct discretization of \eqref{steadystate} \cite{jilkine_phdthesis}.

\subsection{Derivation of the Expression for $K_r$}\label{appendix_Kr}

In this appendix, we derive expressions \eqref{seq} and \eqref{Kreq}.
Consider \eqref{Dinftysteady} when \eqref{cubicmod} is used for the reaction term
where $a>1$. Take any $0\leq K<K_q$ where $K_q$ is given in \eqref{Kq}. Let $v_0$ be the 
middle root of equation \eqref{vequm} and let $u_0=u_m(v_0)$ 
(note that we referred to 
$v_0$ as $v_m^0$ and $u_0$ as $u_m^0$ in Section \ref{otherbif}).
We saw that $(u_0,v_0)$
is a stable spatially homogeneous solution to \eqref{Dinfty} for 
\begin{equation}
\epsilon>\epsilon_0=\frac{1}{\pi\sqrt{(1+(av_0)^2)}}
\end{equation}
where we used \eqref{eps0pf}. At $\epsilon=\epsilon_0$, we demonstrated computationally 
that we have a pitchfork bifurcation. 
We now perform a perturbation calculation to study 
this bifurcation 
(see, for example, \cite{keener_applied_mathematics} 
or \cite{holmes1995introduction}).

Let us restate our problem for future reference. 
For algebraic convenience, we 
shall work with $\lambda=1/\epsilon^2$ instead of $\epsilon$. 
We study the bifurcation of the steady state solution $(u,v)=(u_0,v_0)$ of the system
\begin{align}
\PDD{2}{u}{x}-\lambda(u^2-1)\paren{u+\frac{av}{\sqrt{1+(av)^2}}}&=0,\label{pfstudy1}\\
v+\int_0^1 udx=K, \;\; \at{\PD{u}{x}}{x=0,1}&=0,\label{pfstudy2}
\end{align}
at the bifurcation point $\lambda=\lambda_0$. 
The values $\lambda_0$ and $u_0$ can be expressed in terms of $v_0$:
\begin{equation}
\lambda_0=\pi^2(1+(av_0)^2), \; u_0=\frac{-av_0}{\sqrt{1+(av_0)^2}}\label{lam0u0},
\end{equation}
where $v_0$ is the middle root of:
\begin{equation}
v_0-\frac{av_0}{\sqrt{1+(av_0)^2}}=K.
\end{equation}
Note that $v_0$ can thus be viewed as a function of $K$ where $0\leq K<K_q$.
It is easy to see that $v_0(K)$ is a decreasing function of $K$. As $K$ ranges from 
$0$ to $K_q$, $v_0$ ranges from $0$ to $-\sqrt{a^{2/3}-1}/a$.

We introduce a small parameter $\delta$ and seek a solution of the form:
\begin{equation}
u=u_0+\delta u_1+\delta^2 u_2 +\delta^3 u_3+\cdots.
\end{equation}
We introduce a similar expansion for $\lambda$ and $v$. Substitute these into \eqref{pfstudy1}
and \eqref{pfstudy2}
and collect like terms in $\delta$. The $\mathcal{O}(1)$ relation is just $\lambda_0,u_0,v_0$
substituted into \eqref{pfstudy1} and \eqref{pfstudy2}, 
and thus does not give us anything interesting. 
At $\mathcal{O}(\delta)$ we have:
\begin{equation}
\PDD{2}{u_1}{x}-\lambda_0(u_0^2-1)\paren{u_1+\frac{a}{(1+(av_0)^2)^{3/2}}v_1}=0,\; 
v_1+\int_0^1 u_1 dx=0.\label{Odelta}
\end{equation}
For a function $f$ defined on $0<x<1$, define:
\begin{equation}
\mathcal{Q} f\equiv f-A\int_0^1 fdx, \; A=\frac{a}{(1+(av_0)^2)^{3/2}}.\label{Q}
\end{equation}
Using this and \eqref{lam0u0}, we may rewrite \eqref{Odelta} as:
\begin{equation}
\PDD{2}{u_1}{x}+\pi^2\mathcal{Q}u_1=0,
\end{equation}
where we have Neumann boundary conditions at $x=0,1$.
The only nontrivial solutions to the above are constant multiples of $\cos(\pi x)$.
We thus let:
\begin{equation}
u_1=\cos (\pi x).
\end{equation}
Other choices of $u_1$ merely amounts to a rescaling of $\delta$. 
Note that $v_1=0$ by \eqref{Odelta}.

At $\mathcal{O}(\delta^2)$ we have, after some simplification:
\begin{equation}
\PDD{2}{u_2}{x}+\pi^2\mathcal{Q}u_2=2\lambda_0u_0u_1^2+\lambda_1(u_0^2-1)u_1.
\end{equation}
Given that the operator on the left hand side is self-adjoint with the 
null space spanned by $\cos(\pi x)$, we require that the left hand side 
be orthogonal to this. From this, we easily conclude:
\begin{equation}
\lambda_1=0.
\end{equation}
We may then solve for $u_2$ imposing orthogonality with respect to $u_1=\cos(\pi x)$
to obtain:
\begin{equation}
u_2=\alpha +\beta \cos(\pi x), \; \alpha=\frac{av_0\sqrt{1+(av_0)^2}}{A-1},\; 
\beta=\frac{1}{3}av_0\sqrt{1+(av_0)^2},
\end{equation}
where $A$ was given in \eqref{Q}.

At $\mathcal{O}(\delta^3)$ we have:
\begin{equation}
\PDD{2}{u_3}{x}+\pi^2\mathcal{Q}u_3
=\lambda_2(u_0^2-1)u_1+2\lambda_0u_0u_1\mathcal{Q}u_2+\lambda_0(2u_0u_2+u_1^2)u_1,
\end{equation}
where we have used $\lambda_1=0$. The left hand side must be orthogonal to $u_1$, 
from which we obtain the following expression for $\lambda_2$:
\begin{equation}
\lambda_2=\lambda_0\paren{\int_0^1 (1-u_0^2)u_1^2dx}^{-1}
\paren{\int_0^1 (2u_0u_1^2\mathcal{Q}u_2+(2u_0u_2+u_1^2)u_1^2)dx}. \label{lam2}
\end{equation}
Given that $\lambda=\lambda_0+\delta^2\lambda_2+\cdots$,
the sign of $\lambda_2$ determines whether the pitchfork bifurcation 
is subcritical or supercritical. 
We thus seek the point at which $\lambda_2$ changes sign as $K$ is varied.
The sign of $\lambda_2$ is determined by the sign of the last integral in \eqref{lam2}.
This integral can be computed as:
\begin{equation}
I=\int_0^1 (2u_0u_1^2\mathcal{Q}u_2+(2u_0u_2+u_1^2)u_1^2)dx=u_0\alpha(2-A)+u_0\beta+\frac{3}{8}.
\end{equation}
We may simplify this expression to find:
\begin{equation}
I=(s^2-1)\paren{\frac{5}{3}+\frac{a}{s^3-a}}+\frac{3}{8}, \; s=\sqrt{1+(av_0)^2}.
\end{equation}
Using properties of $v_0(K)$, we see that 
$s$ is an increasing function of $K$ and varies between 
$1\leq s<\sqrt[3]{a}$ for $0\leq K<K_q$. Dividing the above by $s^2-1$, we obtain 
the left hand side of \eqref{seq}, which is monotone in $s$ for $1\leq s< \sqrt[3]{a}$. We see 
that there is only one value of $s$ and hence $K$ at which $I$ changes sign. 
This is the value of $K_r$ we seek.

\subsection*{Acknowledgments}
The authors gratefully acknowledge support from the following sources: 
National Science Foundation (USA) (Grant Number DMS-0914963)  and 
the Alfred P. Sloan Foundation (to YM), 
The Natural Sciences and Engineering Research Council (NSERC), Canada, as well as subcontracts (to LEK) from 
%the National Science Foundation (USA)  (Grant Number DMS-0240770) and 
the National Institutes of Health (Grant Number R01 GM086882)
to Anders Carlsson, Washington University, St Louis. 
We thank A.E. Lindsay for discussions about numerical continuation, 
and A.F.M. Mar\'ee for discussions about Rho GTPase modeling.

\bibliography{Wavepin}
\bibstyle{siam}

\end{document}